\documentclass[aps,prl,twocolumn,superscriptaddress]{revtex4}
\usepackage{graphicx}% Include figure files
\usepackage{dcolumn}% Align table columns on decimal point
\usepackage{bm}% bold math
\usepackage{physics}
\usepackage{amsmath}
\usepackage{mathtools}
\usepackage{amssymb}
\usepackage{array}
\usepackage{color}
\usepackage{xcolor}
\usepackage{nicefrac}
\usepackage[colorlinks=true, breaklinks=true, linkcolor=red, citecolor=blue, urlcolor=blue]{hyperref}

\newcolumntype{P}[1]{>{\centering\arraybackslash}p{#1}}
\newcommand{\bonnpi}{Physikalisches Institut, University of Bonn, Nussallee 12, 53115 Bonn, Germany}
\newcommand{\geneva}{Department of Quantum Matter Physics, University of Geneva, Quai Ernest-Ansermet 24, 1211 Geneva, Switzerland}

\usepackage{color}
\usepackage{ifthen}
\usepackage{booktabs}
\usepackage{multirow}

\begin{document}

\title{Fluctuation-induced Bistability of Fermionic Atoms Coupled to a Dissipative Cavity}

\date{\today}
\author{Luisa Tolle}
\affiliation{\bonnpi}
\author{Ameneh Sheikhan}
\affiliation{\bonnpi}
\author{Thierry Giamarchi}
\affiliation{\geneva}
\author{Corinna Kollath}
\affiliation{\bonnpi}
\author{Catalin-Mihai Halati}
\affiliation{\geneva}

\begin{abstract}
We investigate the steady state phase diagram of fermionic atoms subjected to an optical lattice and coupled to a high finesse optical cavity with photon losses. The coupling between the atoms and the cavity field is induced by a transverse pump beam. Taking fluctuations around the mean-field solutions into account, we find that a transition to a self-organized phase takes place at a critical value of the pump strength. 
In the self-organized phase the cavity field takes a finite expectation value and the atoms show a modulation in the density. 
Surprisingly, at even larger pump strengths two self-organized stable solutions of the cavity field and the atoms occur, signaling the presence of a bistability. We show that the bistable behavior is induced by the atoms-cavity fluctuations and is not captured by the the mean-field approach.   
\end{abstract}
\nopagebreak
{\let\clearpage\relax\maketitle}

The emergence of a bistability is a fascinating effect often occurring in nature. 
Such bistable systems can reach two distinct stable states depending on their history, i.e.~the preparation procedure, or their previous dynamics. 
This concept is important in many scientific fields as for example in biology, chemistry, engineering, and physics. The presence of a bistability leads to profound implications, since the dynamics of the system can become highly sensitive to the initial conditions and external perturbations.

One very well known bistability in physics is the optical bistability \cite{He2014}.
This phenomenon occurs due to nonlinear effects, causing the existence of two stable states with different light intensities. A typical device is the Fabry-Perot cavity containing a nonlinear optical medium with a refractive index depending on the light intensity. Furthermore, the underlying nonlinearity can also emerge from the optomechanical interactions of a cold atomic gas coupled to the field of the optical cavity \cite{RitschEsslinger2013},  e.g.~as it has been recently investigated for the fermionic atoms in optical cavities throughout the BEC-BCS crossover \cite{HelsonBrantut2022}.

Whether bistabilities can also occur in presence of dissipation is of course an important question. 
Bistable behaviors have been indeed identified around dissipative quantum phase transitions, either due to the spontaneous breaking of a weak-symmetry \cite{BaumannEsslinger2010, Keeling_Simons_2010, RitschEsslinger2013, LeBoite_Ciuti_2013, BenitoNavarrete2016, HwangPlenio2018, WilmingEisert2017, Fitzpatrick_Houck_2017, FerreiraRibeiro2019,Stitely_Perkins_2020,Soriente_Zilberberg_2021, Mivehvar_2024}, or due to the first order character of the transition \cite{Labouvie_Ott_2016, BiondiSchmidt2017, Fink_Imamoglu_2017, Foss-Feig_Maghrebi_2017, HannukainenLarson2018, Minganti_Ciuti_2018,Nava_Fabrizio_2019, Garbe_Nori_2020, Ferri_Esslinger_2021, Benary_Ott_2022, Gabor_Domokos_2023}.
In the latter, a hysteresis behavior between the two competing phases can appear, depending on the nature of the initially prepared state.  
However, typically in these cases the bistability occurs within a mean-field approach and the competing states resolve to metastable states when quantum fluctuations around the mean-field solutions are taken into account \cite{Mendoza-Arenas_Jaksch_2016, BenitoNavarrete2016, Landa_Misguich_2020, HalatiKollath2020}.
In such cases the quantum dynamics exhibits a unique steady state, with the density matrix consisting of the admixture of the mean-field bistable states. 
This is in contrast to the presence of multiple stable steady states occurring from the presence of a strong symmetry \cite{BucaProsen2012, AlbertJiang2014,Roberts_Clerk_2020, HalatiKollath2022}.

Intriguingly, in equilibrium, for classical systems or systems described by Ginzburg-Landau type theories, such as magnets, superconductors, or liquid crystals, fluctuations have been shown to change the nature of a phase transition from second-order to a first-order character \cite{Brazovskii1975, Binder1987,HalperinMa1974,HerbutBechhoefer2001, JanoschekPfleiderer2013}, with a coexistence of phases possible around the critical point \cite{HohenbergSwift1995}.

\begin{figure}[h]
    \includegraphics[width=0.48\textwidth]{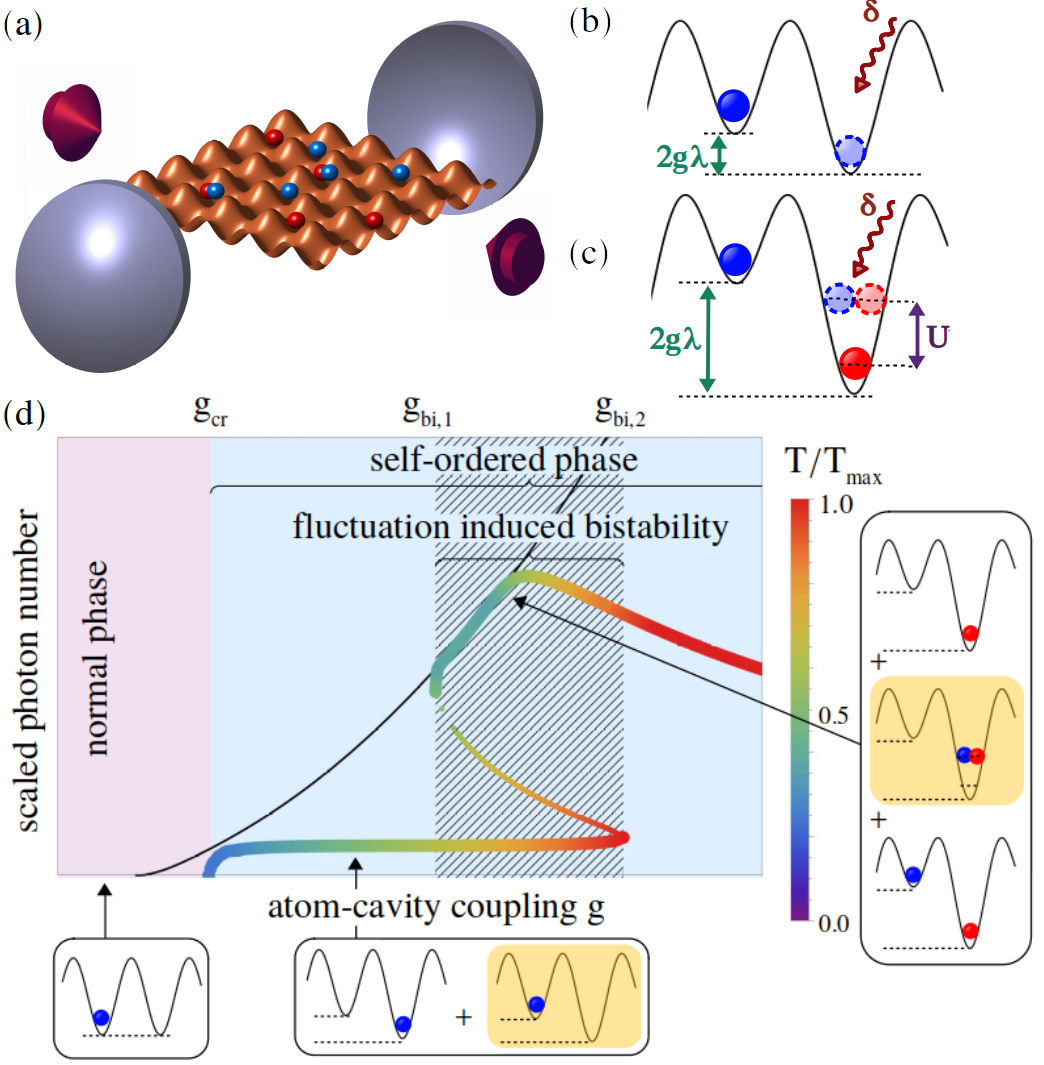} 
    \vspace{-20pt}
    \caption[top: Sketch of interacting fermions on a lattice coupled to a dissipative cavity mode]{(a) Fermi-Hubbard system coupled to a single-mode cavity. The two fermion species have an on-site interaction $U$ and a tunneling amplitude $J$. They are coupled globally to the cavity mode, strength $g$, by pumping with a retro-reflected transverse laser beam. Photons with a detuning $\delta$ leak through the mirrors at rate $\varGamma$. 
    (b),(c) Sketches of important excited-state transitions in the cavity potential with effective energy difference $2g\lambda$, where the rescaled cavity field $\lambda\!=\!\langle\hat{a}+\hat{a}^\dagger\rangle/L^d$ is the order parameter. 
    Resonances between the higher energy sites and doubly occupied sites play a crucial role in the physics (see text). 
    (d) Scaled photon number $N_\text{pho}$ as a function of atom-cavity coupling $g$. The color of the points represents the effective self-consistently determined temperature $T/T_\text{max}$. The black line is the atom-cavity mean-field result ($T\!=\!0$ MF). The sketches highlight the dominant occupations contributing to the different steady states. The yellow-colored contribution is the dominant excited state admixed by the self-consistent effective temperature.}
    \label{fig:cavity_scheme}
\end{figure}

Here we show a novel mechanism, which we refer to as fluctuation-induced bistability, in which bistability occurs in a \emph{dissipative quantum} system, \emph{only} if quantum fluctuations around a mean-field solution are taken into account.
We identify this phenomenon for interacting spinful fermionic atoms confined to optical lattices and coupled to the field of a dissipative cavity, a paradigmatic example of an open quantum system.
We include the fluctuations in the atoms-cavity coupling on top of the mean-field solution resulting from the adiabatic elimination of the cavity field.
The fluctuations induce a thermalization process resulting in steady states with a self-consistently determined temperature.
In the strong atoms-cavity coupling regime two (several) different self-organized states exist, which differ in their effective self-consistently determined temperatures.
Whereas the dominant contribution is the same in both bistable solutions, the nature of the admixed excited states change. 
In particular, in the second steady solution, excited states with double occupied sites become crucial, with a cooling mechanism emerging due to resonant photon-assisted transitions between states with double occupancies and other atomic excitations [see sketch Fig.~\ref{fig:cavity_scheme}(c)].
Thus, we pinpoint the origin of the multistability in the interplay between the \emph{short-range} atomic interactions and the \emph{global-range} coupling via the cavity-induced self-consistent potential. 
We show the existence of the fluctuation-induced bistability is independent of the dimensionality of the atomic gas and we expect that the underlying mechanism is present in a large class of atoms-cavity models with comparable atomic and photonic energy scales. 
The need to consider the fluctuations around the mean-field solution is even more remarkable, due to the general belief of the validity of mean-field methods for long-range couplings. 

The considered platform of ultracold atoms coupled to optical cavities  has established itself for the study of dissipative phenomena \cite{RitschEsslinger2013, MivehvarRitsch2021}. 
Experimentally, bosonic atoms have been placed in cavities both as three-dimensional Bose-Einstein condensates \cite{BaumannEsslinger2010, KroezeLev2018, VaidyaLev2018,KesslerHemmerich2021, FerriEsslinger2021, DreonDonner2022, KongkhambutKessler2022}, or confined additionally to an optical lattice \cite{LandigEsslinger2016,HrubyEsslinger2018, KlinderHemmerich2015, KlinderHemmerich2015b}, more recently also the coupling between cavities and cold fermionic gases has been realized \cite{HelsonBrantut2022, HelsonBrantut2023, ZwettlerBrantut2024, ZhangWu2021, WuWu2023}. 
Generally, these type of systems exhibit a phase transition between a normal phase with an empty cavity and a self-organized phase exhibiting a finite cavity occupation \cite{RitschEsslinger2013}.
The theoretical studies for fermionic systems have focused on 
the nature of the self-organization phase transition \cite{LarsonLewenstein2008b, SunLiu2011, PiazzaStrack2014, PiazzaStrack2014b, ChenZhai2014, KeelingSimons2014, ChenYu2015,Pan2022}, 
employing the attractor dynamics to stabilize exotic states of matter \cite{DongPu2014,  FanJia2018, ColellaChiofalo2018, ColellaRitsch2019,SchlawinJaksch2019, ZhengWang2020,NieZheng2023,ColellaRitsch2019b,MivehvarPiazza2017}, 
topological effects \cite{KollathBrennecke2016, SheikhanKollath2016,SheikhanKollath2016b,Mendez-CordobaQuiroga2020,PanGuo2015,ChandaZakrzewsky2021}, pairing in superfluids \cite{SchlawinJaksch2019b},
or aspects of the non-equilibrium dynamics \cite{WolffKollath2016, MarijanovicDemler2024}.

The dissipative dynamics of the cold atomic gas in an optical cavity, pumped with a standing-wave transverse laser beam far-detuned from the atomic resonance, can be described by a Lindblad master equation
\cite{CarmichaelBook, BreuerPetruccione2002, MaschlerRitsch2008, RitschEsslinger2013, MivehvarRitsch2021}
\begin{equation}
\label{eq:Lindblad_equation_model}
\frac{d}{dt}\hat{\rho}
=-\frac{i}{\hbar}[\hat{H},\hat{\rho}]+ \frac{\varGamma}{2}\left(2\hat{a}\hat{\rho} \hat{a}^\dagger-\hat{a}^\dagger \hat{a}\hat{\rho}-\hat{\rho}\hat{a}^\dagger \hat{a}\right).
\end{equation}
Here $\hat{\rho}$ is the density matrix containing both the atomic and photonic degrees of freedom. The dissipator with amplitude $\varGamma$ describes photon losses from the cavity mode by the photon annihilation jump operator $\hat{a}$. The Hamiltonian is given by $\hat{H}\!=\!\hat{H}_\text{FH}\!+\!\hat{H}_\text{cav}\!+\!\hat{H}_\text{ac}$ \cite{RitschEsslinger2013,MivehvarRitsch2021}. The first term is the Hubbard model
 $\hat{H}_\text{FH}\!=\! -J\!\sum_{\langle j,l\rangle, \sigma}\!\big(\hat{c}_{j\sigma}^\dagger \hat{c}_{l\sigma}\!+\text{H.c.}\big)+ U\!\sum_{j}\!\hat{n}_{j\uparrow}\hat{n}_{j\downarrow}$, 
where $\hat{c}_{j \sigma}$ and $\hat{c}_{j \sigma}^\dagger$ are fermionic operators for the atoms on site $j$ and spin $\sigma\!\in\!\{\uparrow,\downarrow\}$, $\langle j,l\rangle$ denotes neighbouring sites, and the local density operator is $\hat{n}_{j,\sigma}\!=\!\hat{c}_{j,\sigma}^\dagger \hat{c}_{j,\sigma}$. $L$ is the number of sites along each dimension of the lattice system $d$ and $N$ the total number of atoms. $J$ is the tunneling amplitude and $U\!>\!0$ the repulsive on-site interaction strength. 
The second term $\hat{H}_\text{cav}\!=\!\hbar\delta \hat{a}^\dagger\hat{a}$,
describes the cavity mode in the rotating frame of the pump beam, with \textcolor{blue}{$\delta$ the} detuning between the cavity mode and the transverse pump beam. 
The laser-assisted dispersive atoms-cavity coupling, $\hat{H}_\text{ac}\!=\!-\frac{\hbar g}{\sqrt{L^d}}(\hat{a}+\hat{a}^\dagger)\hat{\Delta}$, is chosen such that the cavity mode is commensurable with twice the periodicity of the lattice spacing. 
This effectively creates a bipartite lattice with different sublattices $A,B$ 
and a coupling of the cavity field to the imbalance between the occupation of the different sublattices $\hat{\Delta}\!=\!\sum_{\substack{j\in A,\sigma}}\!\hat{n}_{j\sigma}\!-\!\sum_{\substack{l\in B,\sigma}}\!\hat{n}_{l\sigma}$ with the effective pump strength $g$ \cite{MaschlerRitsch2008}.  
We note that we do not consider in $\hat{H}_\text{ac}$ the dispersive coupling $\propto\hat{a}^\dagger\hat{a}\hat{\Delta}$ \cite{RitschEsslinger2013}, which often leads to bistabilities due to optical non-linearities.

A common method for dealing with cavity-atoms systems is to adiabatically eliminate the cavity mode and perform a mean-field approximation for the coupling term $\hat{H}_\text{ac}$  [in the following called zero-temperature mean-field method ($T\!=\!0$ MF)] \cite{RitschEsslinger2013,MivehvarRitsch2021}.
After a fast decay, the photons are assumed to be in a coherent state with  $\lambda\!\equiv\!\langle \hat{a}+\hat{a}^\dagger\rangle/\sqrt{L^d}\!=\!\frac{2g\delta}{\delta^2+(\varGamma/2)^2}\frac{\langle \hat{\Delta}\rangle}{L^d}$ and the atoms in the self-consistently determined ground state of the effective Hamiltonian
$\hat{H}_\text{eff}\!=\!\hat{H}_\text{FH}\!-\!\hbar g\lambda\hat{\Delta}$.
We note that the MF treatment breaks the weak $\mathbb{Z}_2$ symmetry of the model, $(\hat{a},\hat{\Delta})\!\to\!(-\hat{a},-\hat{\Delta})$, thus, in the following we restrict ourselves to the solution with $\langle \hat{a}\rangle\!\geq\!0$, resulting in the $A$-sublattice to have the lower energy.

However, the fluctuations in the coupling were shown to play an important role in the determination of the steady state in several setups \cite{DamanetKeeling2019, HalatiKollath2020, HalatiKollath2020b, BezvershenkoRosch2021, HalatiKollath2022, JagerBetzholz2022, LinkDaley2022}.
Therefore, we employ a recently developed method, based on the many-body adiabatic elimination technique \cite{Garcia-RipollCirac2009, ReiterSorensen2012, Kessler2012, PolettiKollath2013, SciollaKollath2015, LangeRosch2018}, to include perturbatively the fluctuations on top of the atoms-cavity mean-field \cite{BezvershenkoRosch2021, HalatiKollath2022}.
Due to the complexity of the resulting equations, we perform a further approximation assuming the atomic steady state to be a thermal state, $\textstyle\hat{\rho}_\text{at}\!=\!e^{-\beta\hat{H}_\text{eff}(\lambda)}/Z$. 
This is motivated by the level spacing distribution of $\hat{H}_\text{eff}$ obeying Gaussian orthogonal ensemble (GOE) statistics in the presence of the staggered potential \cite{DeMarcoKollath2022}. 

Importantly, the fluctuations in the atoms-cavity coupling determine the effective temperature at which the atomic system thermalizes and by this causes the admixture of excited states of the effective Hamiltonian in the density matrix of the system, in contrast to the $T\!=\!0$ MF method. The effective inverse temperature $\beta\!=\!1/k_{B} T$ and the cavity field strength $\lambda$ are determined self-consistently from the mean-field relation and the steady state condition of the energy transfer $\pdv{t}\langle\hat{H}_{\text{eff}}\rangle_T\!=\!0$ \cite{BezvershenkoRosch2021, supp}, given by
\begin{equation}
    \pdv{t}\langle\hat{H}_{\text{eff}}\rangle_T\!\propto\!\int\! d\omega \frac{\hbar\omega \Im\left[\chi_T(\omega)\right]}{1-e^{-\beta\hbar\omega}}\frac{\varGamma/(2\pi)}{(\omega+\delta)^2+(\varGamma/2)^2},
\end{equation}
with self-consistently computed retarded susceptibility of the operator $\hat{\Delta}$, $\chi_T(\omega)\!=\!-\frac{i}{\hbar}\int_{0}^{\infty} dt e^{i\omega t}\langle\big[\hat{\Delta}(t),\hat{\Delta}(0)\big]\rangle_T$, which is evaluated for the thermal state $\hat{\rho}_\text{at}$. 
As sketched in Fig.~\ref{fig:cavity_scheme}(d), this approach leads to different values of the temperature throughout the phase diagram and, as we discuss in the following, is one of the crucial ingredients for the occurrence of the bistability.
To gain further insights regarding the self-consistent solutions, it is useful to go to the spectral representation of the energy transfer namely $\pdv{t}\langle\hat{H}_{\text{eff}}\rangle_T\!\propto\!\sum_{n,m} |\Delta_{nm}|^2  \frac{e^{\!-\beta E_{m}} (E_n\!-\!E_m)\varGamma}{\left(E_n\!-\!E_m\!+\!\hbar\delta\right)^2\!+\!\left(\hbar\varGamma/2\right)^2}$, with $E_n$ the energy of the eigenstate $\ket{n}$ of $\hat{H}_\text{eff}$ and $\Delta_{nm}\!=\!\bra{n} \hat{\Delta}\ket{m}$ \cite{BezvershenkoRosch2021, HalatiPhD}.

In the following, we investigate the case of equal number of spin up and down particles. While we do not explicitly consider the total spin symmetry sectors, we checked that we obtain the same effective temperatures in typical spin sectors.

The non-trivial self-consistent solution(s) obtained might not be stable. Thus, we generalized the criterion of their stability \cite{RitschEsslinger2013,Tian2016, HalatiKollath2017} for finite temperature states \cite{supp}. We determine the stability under variation of the cavity field quadratures, considering the $\lambda$-dependence of $E_n(\lambda), \beta(\lambda)$ and $\Delta_{nm}(\lambda)$, and obtain that for stable solutions the following holds 

\begin{align}
    \label{eq:stability}
    \textstyle\frac{\delta^2+\left(\varGamma/2\right)^2}{2\delta g}\!>&\!\textstyle\displaystyle\sum_n\textstyle \frac{\text{e}^{-\beta E_n}}{L^dZ}\!\left[\!\frac{\partial \Delta_{nn}}{\partial \lambda}\!-\Delta_{nn}\!\Big(\!\frac{\partial\beta}{\partial \lambda}E_n\!+\!\beta\frac{\partial E_{n}}{\partial \lambda}\!\Big)\!\right]\nonumber\\
    \textstyle+&\displaystyle\sum_{n,m}\textstyle\frac{\text{e}^{-\beta(E_n+E_m)}}{L^{2d}Z^2}\Delta_{nn}\left(\frac{\partial\beta}{\partial \lambda}E_m+\beta\frac{\partial E_{m}}{\partial \lambda}\right).
\end{align}

\begin{figure}[!hbtp]
    \centering
    \includegraphics[width=0.48\textwidth]{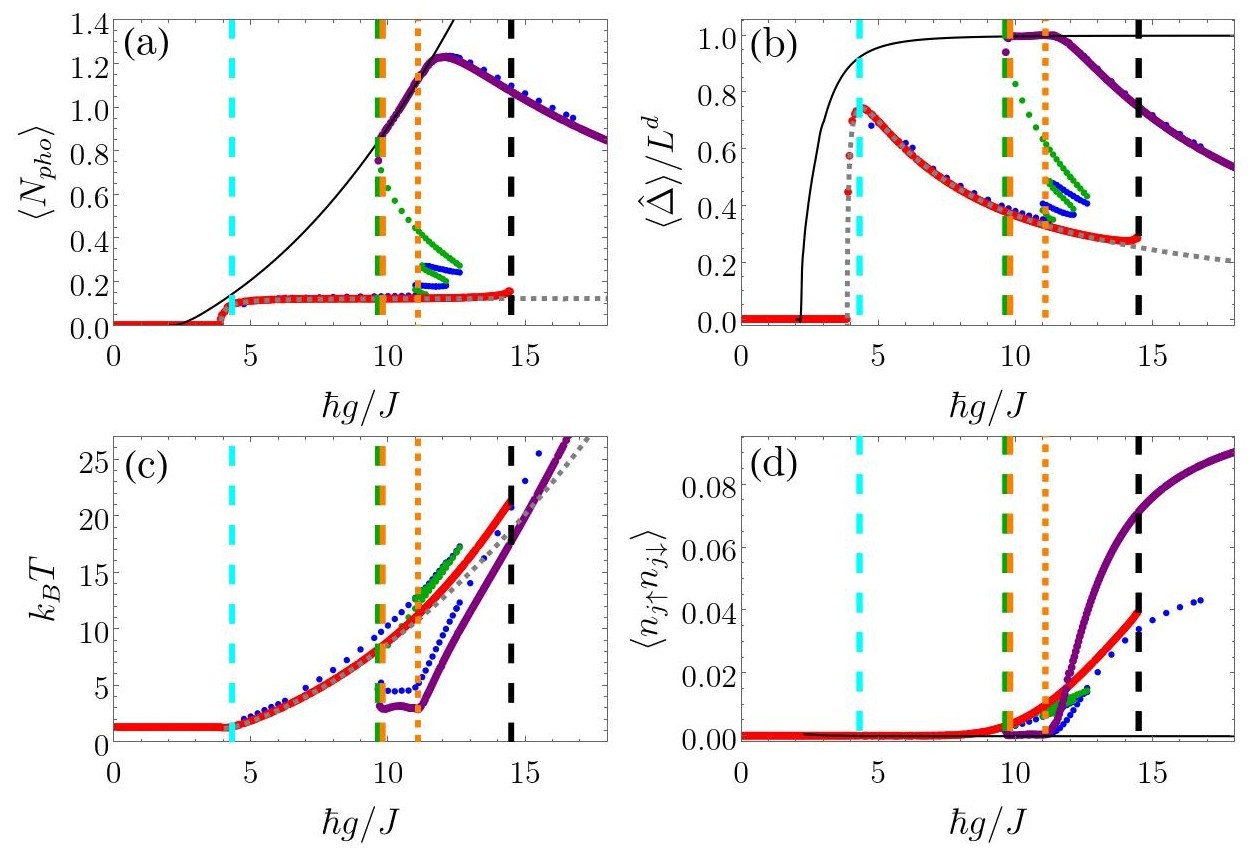}
        \vspace{-20pt}
        \caption{(a) Scaled cavity photon number $\langle N_\text{pho}\rangle\!=\!\langle \hat{a}^\dagger\hat{a}\rangle/L^d\!=\!\frac{\delta^2+(\varGamma/2)^2}{4\delta^2}\lambda^2$, 
        (b) atomic even-odd imbalance $\langle\hat{\Delta}\rangle/L^d$, (c) effective temperature and (d) average double occupancy as a function of the atoms-cavity coupling strength $\hbar g/J$.
        We show the following results: in the thermodynamic limit, $L\!\to\!\infty$ for $U\!\gg\!J$ stable solutions 1 (red) and 2 (purple), and for $U\!\to\!\infty$ (gray, dashed); for $L\!=\!8$ the solution stable (blue), unstable (green); for $L\!=\!8$, within the $T\!=\!0$ MF approximation (black). 
        The other parameters used are $U/J\!=\!40$, $\hbar\delta/J\!=\!5$, $\hbar\varGamma/J\!=\!3$. 
        Vertical lines denote the resonances, $2g\lambda_1\!=\!\delta$ (cyan), $2g\lambda\!=\!U/\hbar$ (black, dashed), $2g\lambda_2\!=\!U/\hbar\!\pm\!\delta$ (orange, short/long-dashed), $4g\lambda_2\!=\!U/\hbar\!-\!\delta$ (green, dashed).}
    \label{fig:cuts_U40_d5_G3}
\end{figure}

In order to obtain the steady state solutions we need to determine the eigenstates of $\hat{H}_\text{eff}$, which we compute by two approaches.
In the numerical approach, we employ the exact diagonalization (ED) method of $\hat{H}_\text{eff}$ in the case of small one-dimensional systems at fixed particle number.
Analytically, we perform a perturbation theory in the kinetic energy $J\!\ll\!U$, which allows us to go to the thermodynamic limit in any dimension.
Here, the atom number is conserved only on average by introducing a chemical potential term in the Hamiltonian, $-\mu\sum_j \hat{n}_j$, and solving $\langle \hat{n}\rangle_T\!\equiv\! N/L^d$ for $\mu$ \cite{supp}. 
The energy transfer in this perturbative approach is given by
\begin{align}
    \label{eq:EOM}
   & \textstyle\frac{\partial}{\partial t}\langle \hat{H}_\text{eff}\rangle_T\textstyle\!\propto\\
    &\qquad\textstyle J^2\bigg[\!\frac{\text{e}^{\beta\mu}+\text{e}^{\beta(3\mu-U)}}{\!2\!g\!\lambda}\Big(\!\frac{-\text{e}^{\beta\hbar\!g\!\lambda}}{(2\!g\!\lambda+\delta)^2+(\varGamma/2)^2}\textstyle+\frac{\text{e}^{-\beta\hbar\!g\!\lambda}}{(2\!g\!\lambda-\delta)^2\!+\!(\varGamma/2)^2}\!\Big)\nonumber\\
    &\quad\displaystyle+\!\sum_{\mathclap{p=\pm 1}}\textstyle\frac{\text{e}^{2\beta\mu}}{2\!g\!\lambda\!-\!pU\!/\!\hbar}\Big(\!\frac{-\text{e}^{\beta(\!2\!\hbar\!g\!\lambda\!-\!pU\!)}}{(2\!g\!\lambda\!-\!pU\!/\!\hbar\!+\!\delta)^2+(\varGamma/2)^2}\!+\!\frac{1}{(2\!g\!\lambda\!-\!pU\!/\!\hbar\!-\!\delta)^2+(\varGamma/2)^2}\!\Big)\!\bigg]\nonumber.
\end{align}

Our results for an atomic filling of $n\!=\!1/2$ show a very rich steady state diagram, see Fig.~\ref{fig:cavity_scheme}(d). 
At low pump power $g$ the system is in the normal phase, with a vanishing photon number. Above a critical value $g_\text{cr}$ the transition to the self-organized phase occurs, signaled by the emergence of a finite cavity field and a corresponding increase of the atomic sublattice density imbalance.
Surprisingly, for even larger pump powers a novel type of bistability,  
occurs within the self-organized phase, which we call \emph{fluctuation-induced bistability}. We emphasize that the bistability is not present in the atoms-cavity mean-field approach [black line, Fig.~\ref{fig:cavity_scheme}(d)] and only occurs determining the effective temperature self-consistently.
We show its absence for an externally fixed finite temperature, see \cite{supp}.

The interplay of the cavity field and the interaction energy of the atoms is crucial for the presence of the bistability. 

The first solution with the lower scaled photon number (marked by subscript $1$) is mainly determined by the balance of the processes originating from transitions between eigenstates of $\hat{H}_\text{eff}$, which are singly occupied on the lower or upper potential sub-lattice, i.e.~$E_n\!-\!E_m\!\approx\! 2\hbar g\lambda_1$, contributing to Eq.~(\ref{eq:EOM}) by the terms in the first line. 
The sign of the terms in Eq.~(\ref{eq:EOM}) implies either a cooling or a heating nature of the processes in the self-consistent dynamics towards the steady state. 
Close to the self-organization threshold $g_\text{cr}$, the resonance between atomic excited states [sketched in Fig.~\ref{fig:cavity_scheme}(d)] and the photon energy leads to a decrease in the self-consistent temperature, see \cite{supp}. For large pump strength in the first solution, to balance the cooling (first term upper line) and heating (second term upper line) the temperature becomes relatively high $k_BT_1\!>\!8 zJ$ [see color coding in Fig.~\ref{fig:cavity_scheme}(d)], with $z$ the lattice coordination number.
In contrast, for the second solution with higher scaled photon number (marked by subscript $2$) the effective temperature is much lower. Transitions between states with either double occupancy, or two singly occupied sites [see Fig.~\ref{fig:cavity_scheme}(c)] captured by the lower line in Eq.~(\ref{eq:EOM}) with $p=1$ become important. The interaction energy is crucial, i.e.~$E_n\!-\!E_m\!\approx\!2\hbar g\lambda_2\!-\!U$.
In particular, the efficient transfer of energy from the atoms to the cavity mode due to the resonances between excited states and the photonic energy $\delta$ leads to a cooling mechanism, driving the atoms to a steady state with a much lower temperature in the bistable region.
The two stable solutions are connected by a third unstable solution [smaller dots in Fig.~\ref{fig:cavity_scheme}(d)]. The occurrence of two stable solutions signals the fluctuation-induced bistability (blue hatched region).

\begin{figure}
    \centering
    \includegraphics[width=0.48\textwidth]{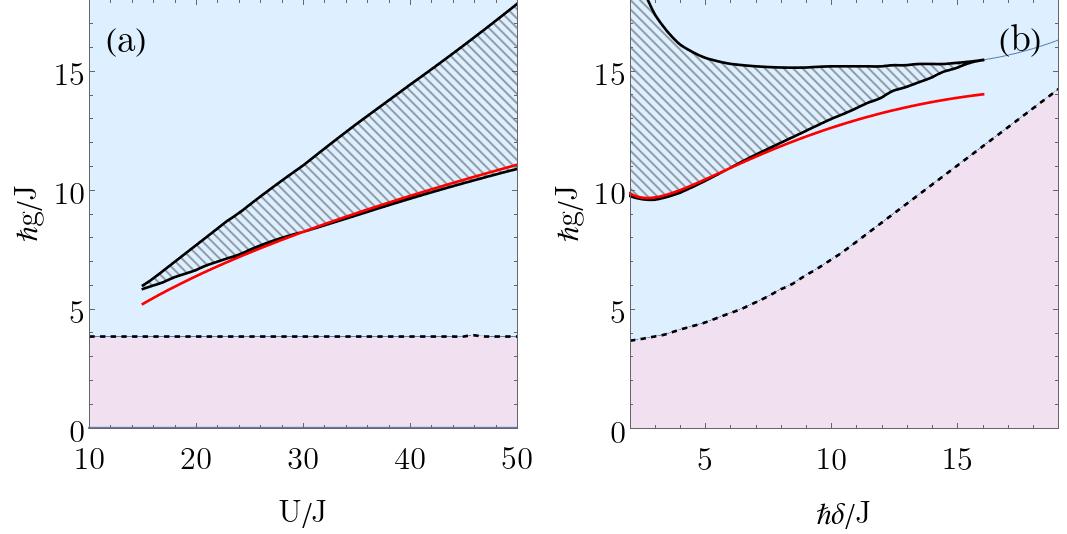}
    \vspace{-20pt}
    \caption{Phase diagrams obtained from solving the equations of motion Eq.~(\ref{eq:EOM}) together with the self-consistency and particle conservation as a function of (a) atoms-field coupling strength $\hbar g/J$ and atomic on-site interaction $U/J$ for $\hbar\delta/J\!=\!5$, $\hbar\varGamma/J\!=\!3$ and (b) $\hbar g/J$ and cavity-pump detuning $\hbar\delta/J$ for $U/J\!=\!40$, $\hbar\varGamma/J\!=\!5$. 
    With purple we mark the normal phase, with blue the self-organized phase, in the hatched region we show the fluctuation-induced bistability region within the self-organized phase. The red line represents the analytic approximation for $\hbar g_{\text{bi},1}/J$, Eq.~(\ref{eq:gbi1_approx}).}
    \label{fig:density_plots_OmegavU_d5_G3}
\end{figure}

To investigate this behavior in more detail, in Fig.~\ref{fig:cuts_U40_d5_G3} we show the results for the photon number, the sublattice imbalance, the atomic temperature, and the double occupancy of the atomic state versus the pump power for the two different methods.
Let us focus first on the perturbative results for small $J$ in the thermodynamic limit (red and purple symbols). The scaled photon number is zero below a critical pump strength, here approximately $\hbar g_\text{cr}/J\!\sim\!3.85$. 
Above $g_\text{cr}$ we see a rapid increase of the cavity field creating an effective staggered potential and therefore a sublattice imbalance which signals the transition to the self-organized phase.
Although such a transition to a self-organized staggered pattern has been known from the $T\!=\!0$ MF approach \cite{RitschEsslinger2013, MivehvarRitsch2021}, the critical pump strength found is considerably lower due to the absence of the fluctuations compared to the value from the finite temperature transition observed here. 
Additionally, a non-monotonous behaviour is observed in the sublattice imbalance for the fermionic atoms [Fig.~\ref{fig:cuts_U40_d5_G3}(b)]. 
In the self-organized regime $\Delta$ has a maximum around $2g\lambda_1\!=\!\delta$ (cyan vertical line) and decreases for larger values of the pump strength. This decrease is mainly due to the quick rise of the finite temperature of the system [Fig.~\ref{fig:cuts_U40_d5_G3}(c)] to a value of the order of the effective staggered potential $2\hbar g\lambda_1$ via the coupling to the cavity.  
At this temperature, reached at intermediate pump strength, $\hbar g/J\gtrsim5$, atoms can be excited to the effectively higher lattice sites leading to the decrease of the sublattice imbalance compared to the atoms-cavity mean-field. This causes a relatively low number of photons per site that only weakly depends on the coupling strength $g$. 
As mentioned, around $\hbar g_{\text{bi},1}/J\!=\!9.65$ a second stable solution appears, with much larger number of photons, an almost maximal value of the atomic density imbalance and a low effective temperature  [Fig.~\ref{fig:cuts_U40_d5_G3}(c)].

We can understand the importance of the 
on-site interaction for the bistability by analyzing Eq.~(\ref{eq:EOM}).
We recover the first solution by neglecting all double occupancy, taking only the first two lines in Eq.~(\ref{eq:EOM}), corresponding to %the limit 
$U\!\to\!\infty$ [gray dashed lines in Fig.~\ref{fig:cuts_U40_d5_G3}].
However, the ending of the first solution and occurrence of the second one can only be obtained by considering terms corresponding to resonances $2g\lambda_2\!=\!U/\hbar\!\pm\!\delta$ \cite{supp}. 
Thus, the bistability is related to processes stemming from the intricate interplay of the short-range atomic interactions and the global coupling to the self-organized photonic field.

For higher values of the pump strength, a drastic rise of the temperature can be found, determining an increase in the occupation of the excited states for the second solution, leading to the presence of double occupancies [Fig.~\ref{fig:cuts_U40_d5_G3}(d)] and to a strong decrease of the scaled photon number and the atomic imbalance [Fig.~\ref{fig:cuts_U40_d5_G3}(a),(b)].
At even larger values of the pump strength $\hbar g_{\text{bi},2}/J\!=\!14.45$ the lower solution and the bistability region end.
So far we focused on the description of the perturbative solution for small hopping $J$ valid in the thermodynamic limit regardless of dimensionality.
However, multistable solutions are observed also using ED to compute the energies and eigenstates of small one-dimensional systems $L\!\leq\!8$ and solve fully Eq.~(\ref{eq:EOM}) (blue, green symbols in Fig.~\ref{fig:cuts_U40_d5_G3}). 
We associate this multi-stable region with the splitting of the resonances by finite size effects and find that the multi-stable region depends strongly on the system size. 
Nevertheless, the finite size systems capture qualitatively the same phenomenon of a multi-stable region caused by resonances, giving us confidence that it is not an artifact of the perturbative approach. 

In Fig.~\ref{fig:density_plots_OmegavU_d5_G3} we show the dependence of the self-organization transition and the fluctuation-induced bistability region on $U/J$ and $\hbar\delta/J$. 
The critical pump strength increases approximately linearly with the detuning $g_\text{cr}\!\sim\!\delta$ at large detuning, similarly to the result obtained for bosonic atoms \cite{BezvershenkoRosch2021}.
As the pump gets further detuned, the effective coupling to the cavity field decreases. Thus, to reach the same effective atom-cavity coupling, a larger amplitude of the pump field is required. 

In contrast, the region of the bistability shows a very strong dependence on the interaction strength.
We can obtain an approximate position of the lower onset \cite{supp}
\begin{equation}
    \label{eq:gbi1_approx}
    g_{\text{bi},1}\!=\!\sqrt{(U/\hbar-\delta)\left[(\varGamma/2)^2+\delta^2\right]/(2\delta)}.
\end{equation}
In particular, the relative strong dependence on $U/J$ is not present in the onset of the self-organization phase for which $g_\text{cr}$ is almost independent on $U/J$ for the considered parameters.
The upper boundary seems to evolve almost linearly with $U/J$, such that a wider bistable region is present at larger interactions. In contrast, increasing $\delta$ causes the bistability region to become smaller and to disappear beyond a certain large value.
Whereas the lower boundary of the bistability region shows a similar rise with $\delta$ than the self-organization transition, the upper boundary becomes almost independent of the detuning above $\varGamma$.

To summarize, we have investigated fermionic atoms in optical lattices coupled to an optical cavity taking fluctuations beyond the mean-field decoupling into account. We find a transition between a normal and a self-organized phase and more surprisingly, the occurrence of a fluctuation-induced bistability region, which does not exist in the $T\!=\!0$ MF solution. 
Within the bistable regime, an efficient cooling of the atomic ensemble takes place by a facilitated energy transfer between excited eigenstates of the atomic system to the photonic mode close to resonances.
Many questions remain open, in particular on the dissipative dynamics of such systems, which might show non-thermal behavior in their approach towards the steady state, caused by the importance of these resonances. 

We expect that the emergence of the fluctuation-induced bistability to be a generic feature of hybrid atoms-cavity systems in which one can control independently the coupling to the cavity and the atomic energies. As in our equations of motion we mainly use the spectrum of effective atomic Hamiltonian, the bistability should emerge for coupled systems where the resonances between the atomic levels and photonic excitations can be obtained.

\emph{Data availability}: The supporting data for this article are openly available at Zenodo \cite{datazenodo}.

\emph{Acknowledgments}:
We thank T.~Donner, J.-P.~Eckmann, S.B.~J\"ager, F.~Mivehvar, H.~Ritsch, A.~Rosch for fruitful discussions.
We acknowledge support by the  Deutsche Forschungsgemeinschaft  (DFG,  German  Research  Foundation) under project number 277625399 - TRR 185 (B4) and project number 277146847 - CRC 1238 (C05) and under Germany’s Excellence Strategy – Cluster of Excellence Matter and Light for Quantum Computing (ML4Q) EXC 2004/1 – 390534769, and by the Swiss National Science Foundation under Division II grant 200020-219400. 
This research was supported in part by the National Science Foundation under Grants No.~NSF PHY-1748958 and PHY-2309135.

\pagebreak
\onecolumngrid

\section{Supplemental Material}

\setcounter{section}{0}
\renewcommand{\thesection}{\Alph{section}}
\setcounter{equation}{0}
\renewcommand{\theequation}{A.\arabic{equation}}

\subsection{Appendix A: Many body adiabatic elimination technique for atoms-cavity coupled systems}
\label{sec:many_body_adiabatic_elimination}

In this section, we outline the derivation of the equation of motion given in the main text [Eq.~(4)]. 
It is derived in the framework of many-body adiabatic elimination taking into account fluctuations on top of the mean-field approximation of the atoms-cavity coupling. 
This mean-field decoupling is typically performed in the approaches involving the adiabatic elimination of the cavity dynamics \cite{RitschEsslinger2013}.

The adiabatic elimination method for many-body quantum systems considers perturbations on top of the decoherence free subspace of the Lindblad master equation  \cite{Garcia-RipollCirac2009, ReiterSorensen2012, Kessler2012, PolettiKollath2013, SciollaKollath2015}.
In our case we employ the many-body adiabatic elimination technique following the mean-field treatment of the atoms-cavity coupling. It allows us to take into account fluctuations in the atoms-cavity coupling perturbatively starting from states obtained after performing a mean-field decoupling of the photonic and atomic degrees of freedom, as we previously developed and benchmarked in Refs.~\cite{BezvershenkoRosch2021, HalatiPhD}. 
We note that our approach does not involve a mean-field treatment of the interaction between the fermionic atoms.

Within the mean-field approximation the steady state of the photons is given by a coherent state, with the field of
\begin{equation}
    \label{eq:selfcon}
    \lambda(\Delta)\equiv\frac{\left\langle\hat{a}+\hat{a}^\dagger\right\rangle}{\sqrt{L^d}}=\frac{2g\delta}{\delta^2+(\varGamma/2)^2}\frac{\langle \hat{\Delta}\rangle}{L^d},
\end{equation}
where $\hat{\Delta}$ is the atomic operator operator to which the cavity field is coupled. In the model we consider in this work, this is given by the sublattice density imbalance
\begin{equation}
    \label{eq:sublattice_imbalance}
    \hat{\Delta}=\displaystyle\sum_{\substack{j\in A,\sigma}}\!\hat{n}_{j\sigma}\!-\!\sum_{\substack{j\in B,\sigma}}\!\hat{n}_{j\sigma}.
\end{equation}
For the atomic part one obtains an effective Hamiltonian which depends on the cavity field
\begin{align}
    \label{eq:Heff_selfcon}
    \hat{H}_\text{eff}= \hat{H}_\text{FH}-\hbar g\!\lambda(\Delta)\hat{\Delta} \quad\text{with }\quad&\hat{H}_{\text{FH}}=-J\!\sum_{\langle j,l\rangle, \sigma}\!\big(\hat{c}_{j\sigma}^\dagger \hat{c}_{l\sigma}+\text{h.c.}\big)+U\displaystyle\sum_{j} \hat{n}_{j\uparrow}\hat{n}_{j\downarrow}-\mu\sum_{j,\sigma}\hat{n}_{j,\sigma}\nonumber.
\end{align}
The coupling $\lambda(\Delta)$ needs to be determined self-consistently using Eq.~(\ref{eq:selfcon}).

However, using this mean-field decoupling any eigenstate of $\hat{H}_\text{eff}$ would be a steady solution and thus, the atomic steady state is not well determined. One common, but somehow arbitrary choice, is to take the ground state of the effective atomic Hamiltonian  \cite{RitschEsslinger2013, MivehvarRitsch2021}, which we refer to as the zero-temperature mean-field approach ($T\!=\!0$ MF) in the main text.
To compute the atomic density matrix one needs to include perturbatively the effects of the fluctuations in the atom-cavity coupling $\delta\hat{H}_\text{ac}\!=\!-\hbar g\big(\frac{\hat{a}+\hat{a}^\dagger}{\sqrt{L^d}}\!-\!\lambda\big)\hat{\Delta}$ \cite{BezvershenkoRosch2021, HalatiPhD}. 

In order to perform the perturbation theory for density matrices within the many-body adiabatic elimination framework we need first to determine the decoherence free and lowest decaying subspaces of the unperturbed Liouvillian $\mathcal{L}_0$. In our case $\mathcal{L}_0\hat{\rho}\!=\!-\frac{i}{\hbar}[\hat{H}_\text{eff}\!+\!\hbar\delta \hat{a}^\dagger \hat{a},\hat{\rho}]\!+\!\frac{\varGamma}{2}\left(2\hat{a}\hat{\rho}\hat{a}^\dagger\!-\!\hat{a}^\dagger \hat{a}\hat{\rho}\!-\!\hat{\rho}\hat{a}^\dagger \hat{a}\right)$, with the perturbation, as mentioned before, defined as $\mathcal{V}(\hat{\rho})\!=\!-\frac{i}{\hbar}[\delta\hat{H}_\text{ac},\hat{\rho}]$. 
The eigenvalues of $\mathcal{L}_0$ have the form $\mathcal{L}_0 \hat{\rho}_\lambda\!=\!(-\lambda^R\!+\!i\lambda^I)\hat{\rho}_\lambda$, where $\lambda^R\!\geq\!0$ and $\lambda^I$ are real numbers, with $\lambda^R$ lying in bands separated by gaps of order $O(\varGamma)$. 
We define $\Lambda_\alpha$ as the subspace of right eigenvectors sharing the same $\lambda^R_\alpha$, with $\Lambda_0$ the decoherence free subspace for which $\lambda^R_0\!=\! 0$ holds. The eigenstates determined by the equation $\mathcal{L}_0 \hat{\rho}^0\!=\!\lambda_0 \hat{\rho}^0$ are given
\begin{equation}
    \begin{aligned}
        \label{eq:ansatz}
        & \hat{\rho}^0= \hat{\rho}_{\text{FH}}^0 \otimes \hat{\rho}_{\text{cav}}^0, \qquad \text{with}\quad\hat{\rho}_{\text{FH}}^0=\ket{n_1(\lambda)}\bra{n_2(\lambda)}, \qquad \hat{\rho}_{\text{cav}}^0=\ket{\alpha(\Delta)}\bra{\alpha(\Delta)},
    \end{aligned}
\end{equation}
where the atomic part is determined by eigenstates $\ket{n(\lambda)}$ of $\hat{H}_\text{eff}$ with energy $E_n(\lambda)$. The cavity state $\hat{\rho}^0_{\text{cav}}$ is given by a coherent state with the field $\alpha(\Delta)/\sqrt{L^d}\!=\!\frac{g}{\delta-i\varGamma/2} \frac{\langle \hat{\Delta}\rangle}{L^d}$.
The eigenstates of $\mathcal{L}_0$ with the lowest decaying rates relevant for us are the following
\begin{equation}
    \begin{aligned}
        \label{eq:eigenstates}
        &\mathcal{L}_0(\ket{n,0}\bra{m,0}) = -i(E_{n}\!-\!E_{m})/\hbar\ket{n,0}\bra{m,0} \\
        & \mathcal{L}_0(\ket{n,1}\bra{m,0}) = [-i((E_{n}\!-\!E_{m})/\hbar+\delta)-\varGamma/2]\ket{n,1}\bra{m,0}, \\
        & \mathcal{L}_0(\ket{n,0}\bra{m,1}) = [-i((E_{n}\!-\!E_{m})/\hbar-\delta)-\varGamma/2]\ket{n,0}\bra{m,1}, 
    \end{aligned}
\end{equation}
where the photonic states are written in the shifted oscillator basis for the cavity operator
\begin{equation}
    \begin{aligned}
        \hat{a} &= \tilde{a}+\alpha, \quad \hat{a}^\dagger = \tilde{a}^\dagger+\alpha^*, \quad (\hat{a}+\hat{a}^\dagger)/\sqrt{L^d} -\lambda= (\tilde{a}+\tilde{a}^\dagger)/\sqrt{L^d},\\
        \ket{0}&\equiv\ket{\alpha}, \quad \tilde{a}\ket{0} =0, \quad \ket{1}\equiv \tilde{a}^\dagger\ket{0}, ~\text{with}~ \bra{0}\ket{1}=0.
    \end{aligned}
\end{equation}
Thus, from Eq.~(\ref{eq:eigenstates}) we can see that the decoherence free subspace is spanned by states of the form $\ket{n,0}\bra{m,0}$ and the excited subspaces by ones that have excitations in the state of the cavity field. A general state in the decoherence free subspace can be written as 
\begin{equation}
    \hat{\rho}^0 = \ket{\alpha(\Delta)}\bra{\alpha(\Delta)} \otimes \hat{\rho}^\text{at}(\lambda), \quad \text{with} \quad \hat{\rho}^\text{at}=\sum_{n_1,n_2}c(n_1,n_2)\ket{n_1(\lambda)}\bra{n_2(\lambda)}.
\end{equation}
States in higher subspaces $\Lambda_{\alpha\neq0}$ decay exponentially in time at a rate $O(\varGamma)$, in our calculation we eliminate adiabatically the contributions stemming from $\Lambda_{1}$, which contain one photonic excitation [see Eq.~(\ref{eq:eigenstates})]. 
In this way one arrives at an equation describing the effective dynamics within the decoherence free subspace \cite{PolettiKollath2013, SciollaKollath2015} which allows us to compute the steady state
\begin{equation}
    \pdv{t}\hat{\rho}^{0} =\mathcal{L}_0\hat{\rho}^{0}+ \frac{1}{\hbar^2}  \hat{P}_0 \left[ \delta\hat{H}_\text{ac},(\mathcal{L}^{\Lambda_1}_0)^{-1} \hat{P}_1 \left[\delta\hat{H}_\text{ac},\hat{\rho}^0\right]\right],
    \label{eq:ae}
\end{equation}
where $\hat{\rho}^0\equiv\hat{\rho}^{\Lambda_0}$ and $\hat{P}_0$ and $\hat{P}_1$ are the projectors to $\Lambda_0$ and $\Lambda_1$, respectively.

Explicitly calculating the steady state of Eq.~(\ref{eq:ae}) using  Eq.~(\ref{eq:eigenstates}) one arrives at an expression where the atomic density matrix is parameterized with a number of coefficients which scales with the square of the atomic Hilbert space dimensions. 
Important to note here is that the resulting equations from perturbative expansion no longer ensure that the solutions are positive semi-definite.
Thus, in order to reduce the complexity of the obtained system of equations and to restrict the solutions to physical density matrices, we make the assumption that we can describe the atomic system with a thermal state \cite{BezvershenkoRosch2021, HalatiPhD}, which is justified if the thermalization time of the atomic system is shorter than the timescale of the scattering from photon fluctuations.
In general, we would have to consider a generalized Gibbs ensemble state for the atoms to take into account all symmetries \cite{HalatiKollath2022, LangeRosch2018}, however, if we consider just a single symmetry sector, or a chaotic system we can describe the atomic density matrix by the thermal state $\hat{\rho}^\text{at}\!\sim\!e^{-\beta\hat{H}_\text{eff}(\lambda)}$.
In our case, for finite values of $\lambda$ the level statistics in each symmetry block of $\hat{H}_\text{eff}$ follows the statistics of Gaussian ensemble of random matrices \cite{DeMarcoKollath2022}, justifying the choice of the ansatz in the self-organized regime.
We note, that in our calculations we do not consider explicitly the total spin quantum number symmetry, however, we numerically checked in each typical symmetry sector we obtain the same temperature.

As for the thermal state we only need to determine a single parameter, $\beta\!=\!1/k_B T$, we use the equation of motion of a single observable, namely the energy transfer, which is given by \cite{BezvershenkoRosch2021, HalatiPhD}
\begin{align}
    \label{eq:thermalss} 
    \frac{\partial}{\partial t}\langle \hat{H}_\text{eff}\rangle_T &=\frac{\hbar^2 g^2\varGamma}{L^dZ}\sum_{n,m}  |\Delta_{m,n}|^2 e^{-\beta E_{m}}\frac{E_n\!-\!E_m}{\left(E_n\!-\!E_m+\hbar\delta\right)^2+\left(\hbar\varGamma/2\right)^2}. \\
    &=\frac{\hbar^2g^2\varGamma}{L^dZ}\sum_{n,m>n}\!|\Delta_{m,n}|^2 (E_n\!-\!E_m)\Bigg[\frac{e^{-\beta E_{m}}}{\left(E_n\!-\!E_m+\hbar\delta\right)^2+\left(\hbar\varGamma/2\right)^2}-\frac{e^{-\beta E_n}}{\left(E_n\!-\!E_m-\hbar\delta\right)^2+\left(\hbar\varGamma/2\right)^2}\Bigg], \nonumber
\end{align}
with $E_n$ the energy of the eigenstate $\ket{n}$ of $\hat{H}_\text{eff}$ and $\Delta_{nm}\!=\!\bra{n} \hat{\Delta}\ket{m}$.

In the steady state, the left hand side vanishes and we only have to solve for $\beta$ by additionally fulfilling the self-consistency condition Eq.~(\ref{eq:selfcon}).

The energy transfer in Lehmann-representation can be rewritten as a function of the retarded susceptibility $\chi(\omega)$.
\begin{equation}
   \label{eq:EOM_susceptibility}
   \frac{\partial}{\partial t}\langle \hat{H}_\text{eff}\rangle_T=\frac{2\hbar g^2}{L^dZ_\text{1D}}\int\! d\omega \frac{\hbar\omega}{1-e^{-\beta\hbar\omega}}\Im\left[\chi_T(\omega)\right]\frac{\varGamma/(2\pi)}{(\omega+\delta)^2+(\varGamma/2)^2},
\end{equation}
where $\chi_T(\omega)=-\frac{i}{\hbar}\int_{0}^{\infty} dt e^{i(\omega+i\epsilon)t}\left\langle\big[\hat{\Delta}(t),\hat{\Delta}(0)\big]\right\rangle_T$.

\setcounter{equation}{0}
\renewcommand{\theequation}{B.\arabic{equation}}

\subsection{Appendix B: Perturbation theory in kinetic terms}
\label{sec:Infinite_size_approximation}

Since that equation of motion given in Eq.~(\ref{eq:thermalss}) is still hard to solve, we describe here how we develop an analytical approach simplifying this equation to reach Eq.~(3) from the main text based on a perturbation theory in the kinetic energy assuming that the hopping amplitude $J$ is the smallest energy scale in the system. The approach allows us to gain a better understanding of the physical processes which cause the fluctuation-induced bistability, as described in the main text. 
Furthermore, this approach is performed in the thermodynamic limit complementing, thus, the direct numerical solution for small systems presented in the main text. 

We split the effective Hamiltonian into an 'unperturbed' part $\hat{H}_0$ and the perturbative kinetic part $\hat{H}_\text{kin}$
 \begin{align}
    \label{eq:Heff_pert}
    \hat{H}_\text{eff}=\hat{H}_{0}+\hat{H}_\text{kin} \quad\text{with }\quad&\hat{H}_{0}=U\displaystyle\sum_{j} \hat{n}_{j\uparrow}\hat{n}_{j\downarrow}-\hbar g\lambda\hat{\Delta}-\mu\sum_{j,\sigma}\hat{n}_{j,\sigma}\\
    &\hat{H}_\text{kin}=\displaystyle-J\!\sum_{\langle j,l\rangle, \sigma}\!\big(\hat{c}_{j\sigma}^\dagger \hat{c}_{l\sigma}+\text{h.c.}\big)\nonumber.
\end{align}
Note, that we added a chemical potential term in order to adjust the correct atomic filling.

We want to rewrite the susceptibility $\chi_T(\omega)$ in terms of the perturbation $\hat{H}_\text{kin}$. To do this, we first compute the equation of motion of the operator $\hat{\Delta}$. Since $\hat{\Delta}$ consists of local densities all terms of $\hat{H}_\text{eff}$ except for the kinetic part commute with it and we obtain
\begin{align}
    \frac{\partial}{\partial t}\hat{\Delta}(t)&=\frac{\partial}{\partial t}e^{i\hat{H}_\text{eff}t/\hbar}\hat{\Delta}e^{-i\hat{H}_\text{eff}t/\hbar}=-\frac{i}{\hbar}e^{i\hat{H}_\text{eff}t/\hbar}\big[\hat{H}_\text{eff},\hat{\Delta}\big]e^{-i\hat{H}_\text{eff}t/\hbar}\\
    &=-\frac{i}{\hbar}e^{i\hat{H}_0t/\hbar}\big[\hat{H}_\text{kin},\hat{\Delta}\big]e^{-i\hat{H}_0t/\hbar}=-\frac{i}{\hbar}\big[\hat{H}_\text{kin},\hat{\Delta}\big]_t.\nonumber
\end{align}
Using this result we have
\begin{align}
    \label{eq:Susceptibility}
    \chi_T(\omega)&=-\frac{i}{\hbar}\int_{0}^{\infty} dt e^{i(\omega+i\epsilon)t}\left\langle\big[\hat{\Delta}(t),\hat{\Delta}(0)\big]\right\rangle_T=-\frac{1}{\hbar\omega}\int_{0}^{\infty} dt\frac{\partial}{\partial t}\big(e^{i(\omega+i\epsilon)t}\big)\left\langle\big[\hat{\Delta}(t),\hat{\Delta}(0)\big]\right\rangle_T\\
    &\stackrel{P.I.}{=}-\frac{1}{\hbar\omega}\left[e^{i(\omega+i\epsilon)t}\left\langle\big[\hat{\Delta}(t),\hat{\Delta}(0)\big]\right\rangle_T\right]_0^\infty+\frac{1}{\hbar\omega}\int_{0}^{\infty} dt e^{i(\omega+i\epsilon)t}\frac{\partial}{\partial t}\left\langle\big[\hat{\Delta}(t),\hat{\Delta}(0)\big]\right\rangle_T\nonumber\\
    &=-\frac{i}{\hbar^2\omega}\int_{0}^{\infty} dt e^{i(\omega+i\epsilon)t}\left\langle\Big[\big[\hat{H}_\text{kin},\hat{\Delta}\big]_t,\hat{\Delta}(0)\Big]\right\rangle_T=-\frac{1}{\hbar^2\omega^2}\int_{0}^{\infty} dt \frac{\partial}{\partial t}\big( e^{i(\omega+i\epsilon)t}\big)\left\langle\Big[\big[\hat{H}_\text{kin},\hat{\Delta}\big]_t,\hat{\Delta}\Big]_{0}\right\rangle_T\nonumber\\
    &\stackrel{P.I.}{=}-\frac{i}{\hbar^3\omega^2}\int_{0}^{\infty} dt e^{i(\omega+i\epsilon)t}\left\langle\Big[\big[\hat{H}_\text{kin},\hat{\Delta}\big]_t,\big[\hat{H}_\text{kin},\hat{\Delta}\Big]_{0}\Big]\right\rangle_T\nonumber.
\end{align}
In this calculation we used the time-translational symmetry $\big[\hat{\Delta}(t),\hat{\Delta}(0)\big]\!=\!\big[\hat{\Delta}(0),\hat{\Delta}(-t)\big]$ and we performed partial integration (P.I.).
The double commutator expression of the susceptibility at which we arrived provides a nice way to perform the perturbation theory.
By computing the expectation values with respect to the unperturbed Hamiltonian $\hat{H}_0$, defined as $\langle...\rangle_{T,\hat{H}_0}=\Tr\big[e^{-\beta\hat{H}_0}...\big]$, we obtain perturbative result in the kinetic terms on the order of $\mathcal{O}(J^2)$. 
The explicit calculation of the commutators and expectation value results in the following 

\begin{align}
    \big[\hat{H}_\text{kin},\hat{\Delta}\big]_t&=e^{i\hat{H}_0 t/\hbar}\big[\hat{H}_\text{kin},\hat{\Delta}\big]_{0}e^{-i\hat{H}_0 t/\hbar}=Je^{i\hat{H}_0 t/\hbar}\Big(\!\sum_{\langle j,l\rangle,\alpha}\!\text{sgn}(l\!-\!j)\hat{c}^\dagger_{j\alpha}\hat{c}_{l\alpha}\Big)e^{-i\hat{H}_0 t/\hbar}.
\end{align}
We determine the time evolution of the operators by considering the action on the local basis states $\{\ket{0},\ket{\uparrow},\ket{\downarrow},\ket{\uparrow\downarrow}\}$ and obtain
\begin{equation}
    \hat{c}_{j\alpha}(t)=e^{it(g\lambda(-1)^j+\mu/\hbar)}\hat{c}_{j\alpha}\big(1-\hat{n}_{j\bar{\alpha}}+e^{iUt/\hbar}\hat{n}_{j\bar{\alpha}}\big).
\end{equation}
Thus, we have
\begin{align}
    \big[\hat{H}_\text{kin},\hat{\Delta}\big]_t=-J\sum_{\langle j,l\rangle,\alpha}\!\text{sgn}(l\!-\!j)e^{-2ig\lambda(-1)^jt}\hat{c}^\dagger_{j\alpha}\hat{c}_{l\alpha}\big(&1-\hat{n}_{j\bar{\alpha}}-\hat{n}_{l\bar{\alpha}}+2\hat{n}_{j\bar{\alpha}}\hat{n}_{l\bar{\alpha}}\\&+e^{iUt/\hbar}\hat{n}_{j\bar{\alpha}}(1-\hat{n}_{l\bar{\alpha}})+e^{-iUt/\hbar}\hat{n}_{l\bar{\alpha}}(1-\hat{n}_{j\bar{\alpha}})\big)\nonumber
\end{align}
The Hamiltonian $\hat{H}_0$ is diagonal in the basis of local densities. Thus, when we compute the expectation value of the commutator $\left\langle\Big[\big[\hat{H}_\text{kin},\hat{\Delta}\big]_t,\big[\hat{H}_\text{kin},\hat{\Delta}\big]_{0}\Big]\right\rangle_{T,\hat{H}_0}$ we only obtain non-vanishing contributions when we return to the initial density distribution.

We note that in arbitrary dimensions, as we consider only terms up to $\mathcal{O}(J^2)$, we can do the following decomposition
\begin{equation}
    \label{eq:direction_decomposition}
    \left\langle\Big[\big[\hat{H}_\text{kin},\hat{\Delta}\big]_t,\big[\hat{H}_\text{kin},\hat{\Delta}\big]_{0}\Big]\right\rangle_{T,\hat{H}_0}\!=\!\sum_s\left\langle\Big[\big[\hat{H}^s_\text{kin},\hat{\Delta}\big]_t,\big[\hat{H}^s_\text{kin},\hat{\Delta}\big]_{0}\Big]\right\rangle_{T,\hat{H}_0},
\end{equation}
where $\hat{H}_\text{kin}$ is split into terms along the direction $s\!=\!x,y,z$ of the hopping. The expectation value is given by
\begin{align}
\label{eq:HkinOHkinO}
    \left\langle\Big[\big[\hat{H}_\text{kin},\hat{\Delta}\big]_t,\big[\hat{H}_\text{kin},\hat{\Delta}\big]_{0}\Big]\right\rangle_{T,\hat{H}_0}\!=-\frac{2iJ^2}{Z}\sum_{\substack{\langle j,l\rangle,\alpha\\j<l}}&\sin(2g\lambda(-1)^jt)\left\langle(1-\hat{n}_{j\bar{\alpha}}-\hat{n}_{l\bar{\alpha}}+2\hat{n}_{j\bar{\alpha}}\hat{n}_{l\bar{\alpha}})(\hat{n}_{j\alpha}-\hat{n}_{l\alpha})\right\rangle_{T,\hat{H}_0}\nonumber\\
    &+\sin((2g\lambda(-1)^j-U/\hbar)t)\left\langle\hat{n}_{j\bar{\alpha}}(1-\hat{n}_{l\bar{\alpha}})(\hat{n}_{j\alpha}-\hat{n}_{l\alpha})\right\rangle_{T,\hat{H}_0}\nonumber\\
    &+\sin((2g\lambda(-1)^j+U/\hbar)t)\left\langle\hat{n}_{l\bar{\alpha}}(1-\hat{n}_{j\bar{\alpha}})(\hat{n}_{j\alpha}-\hat{n}_{l\alpha})\right\rangle_{T,\hat{H}_0}
\end{align}
In the next section we explicitly compute the susceptibility for the case of a one-dimensional atomic system and afterwards argue that we obtain the same also for two-dimensions.

\subsubsection{1.~Results for a one-dimensional atomic system}
\label{subsec:infinite_size_approximation_1D}

In one-dimension in the thermodynamic limit, by considering the kinetic terms to the order of $\mathcal{O}(J^2)$, we effectively reduce the system to a two-site unit cell. 
Thus, we can compute the expectation values as
\begin{align}
    \left\langle\Big[\big[\hat{H}_\text{kin},\hat{\Delta}\big]_t,\big[\hat{H}_\text{kin},\hat{\Delta}\big]_{0}\Big]\right\rangle_{T,\hat{H}_0}\!=\frac{16iJ^2}{Z_\text{1D}}\Big[&2\sin(2g\lambda t)\big(e^{\beta\mu}+e^{\beta(3\mu-U)}\big)\sinh(\beta\hbar g\lambda)\\
    &-\sin((2g\lambda-U/\hbar)t)e^{2\beta\mu}\big(1-e^{\beta(2\hbar g\lambda-U)}\big)\nonumber\\
    &+\sin((2g\lambda+U/\hbar)t)e^{2\beta\mu}\big(1-e^{-\beta(2\hbar g\lambda+U)}\big)\Big]\nonumber.
\end{align}
Inserting this result into Eq.~(\ref{eq:Susceptibility}) we have
\begin{align}
    \chi_T(\omega)=\frac{16J^2}{\hbar^3\omega^2Z_\text{1D}}\int_{0}^{\infty} dt e^{i(\omega-i\epsilon)t}\Big[&2\sin(2g\lambda t)\big(e^{\beta\mu}+e^{\beta(3\mu-U)}\big)\sinh(\beta\hbar g\lambda)\\
    &-\sin((2g\lambda-U/\hbar)t)e^{2\beta\mu}\big(1-e^{\beta(2\hbar g\lambda-U)}\big)\nonumber\\
    &+\sin((2g\lambda+U/\hbar)t)e^{2\beta\mu}\big(1-e^{-\beta(2\hbar g\lambda+U)}\big)\Big]\nonumber\\
    =\frac{16J^2}{\hbar^3\omega^2Z_\text{1D}}\bigg[&\frac{2g\lambda}{(\omega+i\epsilon)^2-(2g\lambda)^2}\big(e^{\beta\mu}+e^{\beta(3\mu-U)}\big)\sinh(\beta\hbar g\lambda)\nonumber\\
    &-\frac{(2g\lambda-U/\hbar)}{(\omega+i\epsilon)^2-(2g\lambda-U/\hbar)^2}e^{2\beta\mu}\big(1-e^{\beta(2\hbar g\lambda-U)}\big)\nonumber\\
    &+\frac{(2g\lambda+U/\hbar)}{(\omega+i\epsilon)^2-(2g\lambda+U/\hbar)^2}e^{2\beta\mu}\big(1-e^{-\beta(2\hbar g\lambda+U)}\big)\bigg]\nonumber
\end{align}
We can now compute the time dependence of the effective Hamiltonian, Eq.~(\ref{eq:EOM_susceptibility}), for this perturbative approach, which we discuss in the main text,
\begin{subequations}
\label{eq:system_of_equations_1D}
\begin{align}
    \label{eq:EOM_1D}
    \frac{\partial}{\partial t}\langle \hat{H}_\text{eff}\rangle_T&=\frac{2\hbar g^2}{Z_{\text{1D}}}\int d\omega \frac{\hbar\omega}{1-e^{-\beta\hbar\omega}}\Im\left[\chi_T(\omega)\right]\frac{\varGamma/(2\pi)}{(\omega+\delta)^2+(\varGamma/2)^2}\\
    &=\frac{8\hbar J^2g^2\varGamma}{Z_{\text{1D}}}\Bigg[\left(\frac{-\exp(\beta \hbar g\lambda)}{(2\hbar g\lambda+\hbar\delta)^2+(\hbar\varGamma/2)^2}+\frac{\exp(-\beta\hbar g\lambda)}{(2\hbar g\lambda-\hbar\delta)^2+(\hbar\varGamma/2)^2}\right)\frac{e^{\beta\mu}+e^{\beta(3\mu-U)}}{2\hbar g\lambda}\label{eq:EOM_1D_terms}\\
    &+\left(\frac{-\exp[\beta(2\hbar g\lambda-U)]}{(2\hbar g\lambda-U+\hbar\delta)^2+(\hbar\varGamma/2)^2}+\frac{1}{(2\hbar g\lambda-U-\hbar\delta)^2+(\hbar\varGamma/2)^2}\right)\frac{e^{2\beta\mu}}{2\hbar g\lambda-U}\nonumber\\
    &+\left(\frac{-\exp[\beta(2\hbar g\lambda+U)]}{(2\hbar g\lambda+U+\hbar\delta)^2+(\hbar\varGamma/2)^2}+\frac{1}{(2\hbar g\lambda+U-\hbar\delta)^2+(\hbar\varGamma/2)^2}\right)\frac{e^{2\beta\mu}}{2\hbar g\lambda+U}\Bigg],\nonumber
\end{align}
where we used $\frac{A}{(\omega+i\epsilon)^2-A^2}=\frac{1}{2}\big(\frac{1}{\omega+i\epsilon-A}-\frac{1}{\omega+i\epsilon+A}\big)$ and $\lim_{\epsilon\to 0^+}\left[\Im\left[\frac{1}{x+i \epsilon}\right]\right]=-\pi\delta(x)$.

Similarly, in the local basis we can calculate the mean-field self consistency condition and the total particle-number, which allows to determine the chemical potential $\mu$ required to fix the particle density to the desired filling.
\begin{align}
    \label{eq:pc_1D}
    \left\langle \hat{n}\right\rangle_T&= \frac{1}{Z_{\text{1D}}}\big(\!\cosh(\beta\hbar g\lambda)\big(e^{\beta\mu}+e^{\beta(3\mu-U)}\big)+e^{2\beta\mu}\left(2+e^{-\beta U}\!\cosh(2\beta\hbar g\lambda)\right)+e^{\beta(4\mu-2U)}\big)\\
    \lambda&=\frac{2g\delta}{\delta^2+(\varGamma/2)^2}\frac{\langle\hat{\Delta}\rangle_T}{L^d}=\frac{1}{Z_{\text{1D}}}\frac{4g\delta}{\delta^2+(\varGamma/2)^2}\big(\sinh(\beta\hbar g\lambda)\big(e^{\beta\mu}+e^{\beta(3\mu-U)}\big)+e^{\beta(2\mu-U)}\sinh(2\beta\hbar g\lambda)\big)\label{eq:sc_1D},
\end{align}
with the partition function (scaled with the system size) given by
\begin{align}
    \label{eq:partition_function_1D}
    Z_{\text{1D}}=&1+4\cosh(\beta\hbar g\lambda)\big(e^{\beta\mu}+e^{\beta(3\mu-U)}\big)+2e^{2\beta\mu}\left(2+e^{-\beta U}\!\cosh(2\beta\hbar g\lambda)\right)+e^{\beta(4\mu-2U)}.
\end{align}
\end{subequations}
In order to determine the steady state solutions and obtain the results presented in the main text, we need to solve the coupled self-consistent equations Eqs.~(\ref{eq:system_of_equations_1D}).

\subsubsection{2.~The case of higher dimensional atomic lattices}
\label{subsec:infinite_size_approximation_higher_dimension}

As in the results from the previous section we consider the kinetic processes up to the order $\mathcal{O}(J^2)$ we expect to obtain similar results also in the case in which the atoms are confined to a higher dimensional lattice. 
This is motivated by the fact that in Eq.~(\ref{eq:direction_decomposition}) we can decompose the quantity entering the calculation of the susceptibility $\chi_T(\omega)$ to independent contributions stemming from each direction of the lattice. Processes in which the particles tunnel in multiple lattice directions are contained only at higher orders. Thus, the higher dimensional susceptibility is different to the one-dimensional susceptibility only by a multiplicative factor, which can be thought as a rescaled tunneling amplitude.
As we are only interested in the steady state solution of Eq.~(\ref{eq:EOM_1D}), we obtain the same equations regardless of the dimension in which we confine the fermionic atoms.
We note that the rates of approaching the steady state depend on the dimensionality and coordination number of the lattice.

We checked this result explicitly for the case of a two-dimensional square lattice. In this case one needs to consider a $2\times 2$-site unit cell in Eq.~(\ref{eq:HkinOHkinO}) and, by noting that the partition function is $Z_{\text{2D}}\!=\!Z_\text{1D}^2$, we recovered the same equations as for the one-dimensional case Eqs.~(\ref{eq:system_of_equations_1D}).

\setcounter{equation}{0}
\renewcommand{\theequation}{C.\arabic{equation}}

\subsection{Appendix C: Cooling resonance close to self-ordering transition}

\begin{figure}[h]
    \includegraphics[width=0.45\textwidth]{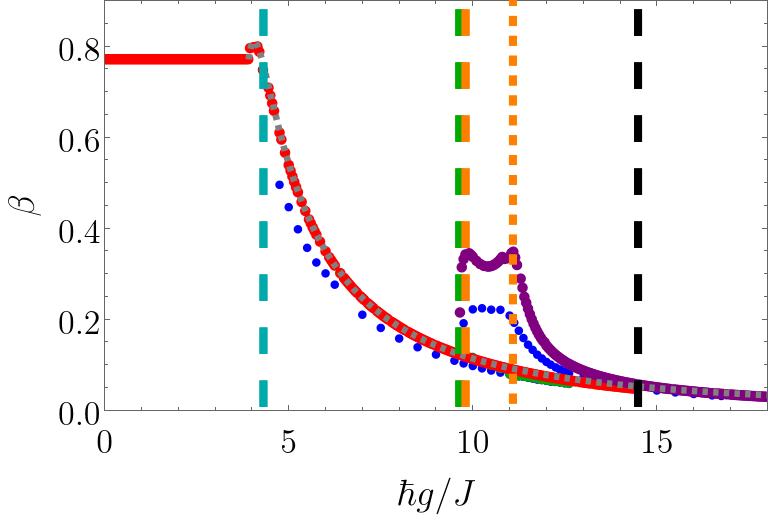} 
    \caption[beta vs g]{Inverse effective temperature $\beta=1/k_BT$ as a function of atom-cavity coupling $\hbar g/J$.  We show the following results: in the thermodynamic limit, $L\!\to\!\infty$ for $U\!\gg\!J$ stable solutions 1 (red) and 2 (purple), and for $U\!\to\!\infty$ (gray, dashed); for $L\!=\!8$ the solution stable (blue), unstable (green). 
    The other parameters used are $U/J\!=\!40$, $\hbar\delta/J\!=\!5$, $\hbar\varGamma/J\!=\!3$. 
    Vertical lines denote the resonances, $2g\lambda_1\!=\!\delta$ (cyan), $2g\lambda\!=\!U/\hbar$ (black, dashed), $2g\lambda_2\!=\!U/\hbar\!\pm\!\delta$ (orange, short/long-dashed), $4g\lambda_2\!=\!U/\hbar\!-\!\delta$ (green, dashed).}
    \label{fig:betavsg}
\end{figure}

In order to highlight the minima occurring in the effective self-consistent temperature discussed in the main text, we plot in Fig.~\ref{fig:betavsg} the inverse temperature $\beta=1/k_BT$ corresponding to the temperature data shown in Fig.~2(c) of the main text.
In particular, we observe that besides the bistable regime, also close to the self-organization threshold $g_\text{cr}$, the resonance between atomic excited states and the photon energy leads to a decrease in the self-consistent temperature.

\setcounter{equation}{0}
\renewcommand{\theequation}{D.\arabic{equation}}

\subsection{Appendix D: Solution at fixed finite temperature}

\begin{figure}[h]
    \includegraphics[width=0.49\textwidth]{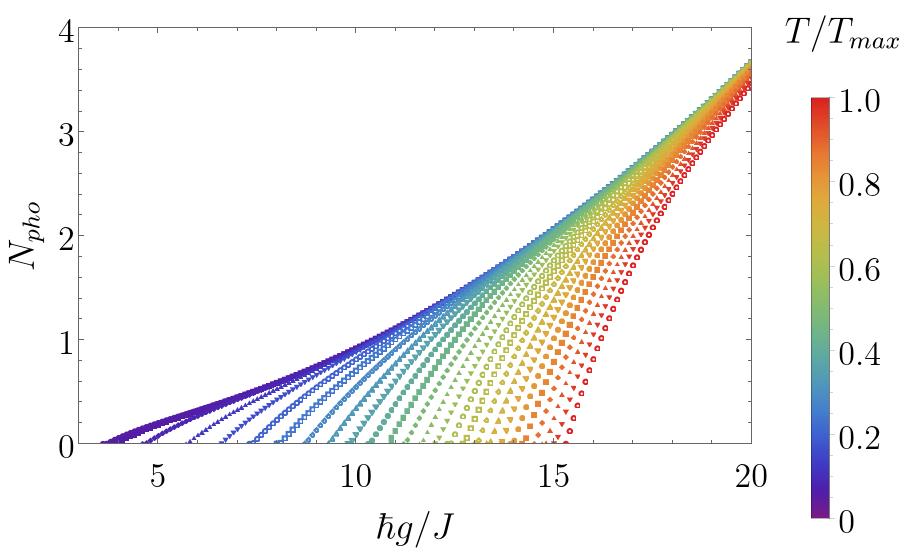} 
    \caption[Npho at fixed finite temperature]{Scaled photon number $N_\text{pho}$ as a function of atom-cavity coupling $\hbar g/J$. The color of the points represent the fixed effective rescaled temperature $T/T_\text{max}$. The parameters used are $U/J\!=\!40$, $\hbar\delta/J\!=\!5$, $\hbar\varGamma/J\!=\!3$, with $k_BT_\text{max}/J\!=\!25$.}
    \label{fig:fixed_temp}
\end{figure}

Solving Eqs.~(\ref{eq:system_of_equations_1D}) at fixed finite temperature values leads to a unique solution for the cavity field parameter $\lambda$ and the corresponding field occupation for each value of $\beta$, covering the entire range obtained by solving for both $\beta$ and $\lambda$. Results are shown in Fig.~\ref{fig:fixed_temp} for the parameters chosen in the main text of the paper. For the fixed temperature for each value of the transverse pump strength only one solution exists. For the self-consistently determined temperature, the bistability occurs within this parameter regime, showing that the bistability is caused by the self-consistently determined temperature.

\setcounter{equation}{0}
\renewcommand{\theequation}{E.\arabic{equation}}

\subsection{Appendix E: Stability condition for thermal steady states}
\label{sec:stability_condition}

In order to determine which solutions from our self-consistent equations of motion, Eqs.~(\ref{eq:system_of_equations_1D}), are stable, we check the stability against small perturbations in the cavity field. 
This has been previously done for atoms-cavity systems in the case in which the atoms are described by a pure state, e.g.~in Refs.~\cite{Tian2016, HalatiKollath2017}. In the following, we extend the derivation for the situation in which the atomic state is given by a density matrix, in particular, by a thermal state. 
To determine the stability, we consider the dependence of the eigenenergies of the atomic effective Hamiltonian $E_n(\lambda)$, the inverse temperature $\beta(\lambda)$ and matrix elements of the imbalance operator $\Delta_{nm}(\lambda)$ on the cavity field $\lambda$.
Note, that all expectation values are calculated w.r.t the self-consistently determined thermal steady states. For readability we omit the subindex used until here $\langle...\rangle_T=\langle...\rangle$.
We compute the equations of motion for the quadratures of the cavity field, defined as
\begin{equation}
    \lambda=\langle\hat{a}^\dag\!+\!\hat{a}\rangle/\sqrt{L^d},\qquad p=i\langle \hat{a}^\dag\!-\!\hat{a}\rangle/\sqrt{L^d},
\end{equation}
from the dynamics of $\hat{a}$, given by
\begin{equation}
    \frac{d}{dt}\langle\hat{a}\rangle=-i\delta\langle \hat{a}\rangle-\frac{\varGamma}{2}\langle\hat{a}\rangle+i\frac{g}{\sqrt{L^d}}\langle \hat{\Delta} \rangle^{(s)},
\end{equation}
to obtain
\begin{equation}
    \frac{d\lambda}{dt}=-\frac{\varGamma}{2}\lambda+\delta p,\qquad\frac{dp}{dt}=-\delta\lambda-\frac{\varGamma}{2} p+\frac{2g}{L^d}\langle \hat{\Delta}\rangle^{(s)},
\end{equation}
where the expectation value $\langle ...\rangle^{(s)}$ is computed with respect to the self-consistent thermal density matrix.

In the next step we find the stationary solutions of these equations
\begin{equation}
    \lambda^{(s)}=\frac{2\delta g}{\delta^2+(\varGamma/2)^2}\frac{\langle \hat{\Delta}\rangle}{L^d}^{(s)},\qquad p^{(s)}=\frac{\varGamma g}{\delta^2+(\varGamma/2)^2}\frac{\langle \hat{\Delta}\rangle}{L^d}^{(s)}.
\end{equation}
To study the stability of the stationary solutions we consider linear fluctuations around the stationary values
\begin{equation}
    \lambda=\lambda^{(s)}+\tilde{\lambda},\qquad
    p=p^{(s)}+\tilde{p}, \qquad\langle \hat{\Delta}\rangle=\langle \hat{\Delta}\rangle^{(s)}+\frac{\differential \langle \hat{\Delta}\rangle^{(s)}}{\differential \lambda^{(s)}}\tilde{\lambda}.
\end{equation}
As the expectation value of the atomic imbalance is given by a thermal state $\langle\hat{\Delta}\rangle^{(s)}\!=\!\frac{1}{Z(\lambda)}\sum_n\!\text{e}^{-\beta(\lambda)E_n(\lambda)}\Delta_{nn}$, its differential will be given by
\begin{align}
    \frac{\differential \langle \hat{\Delta}\rangle^{(s)}}{\differential \lambda^{(s)}}&=\frac{1}{Z(\lambda)^2}\left\{\sum_n\frac{\partial}{\partial \lambda}\left[e^{-\beta(\lambda)E_n(\lambda)}\Delta_{nn}(\lambda)\right]Z(\lambda) +\left[e^{-\beta(\lambda)E_n(\lambda)}\Delta_{nn}(\lambda)\right]\frac{\partial}{\partial \lambda}Z(\lambda)\right\}\\
    &=\frac{1}{Z}\sum_n e^{-\beta E_n}\left[\frac{\partial \Delta_{nn}}{\partial \lambda}-\Delta_{nn}\left(\frac{\partial \beta}{\partial \lambda}E_n+\beta\frac{\partial E_{n}}{\partial \lambda}\right)\right]+\frac{1}{Z^2}\sum_{n,m} e^{-\beta (E_n+E_m)}\Delta_{nn}\left(\frac{\partial \beta}{\partial \lambda}E_m+\beta\frac{\partial E_{m}}{\partial \lambda}\right)\nonumber.
\end{align}
Thus, the dynamics of the fluctuations is given by 
\begin{equation}
    \frac{d\tilde{\lambda}}{d t}=-\frac{\varGamma}{2}\tilde{\lambda}+\delta \tilde{p},\qquad\frac{d\tilde{p}}{d t}=-\delta\lambda-\frac{\varGamma}{2} \tilde{p}+\frac{2g}{L^d}\frac{\differential \langle \hat{\Delta}\rangle^{(s)}}{\differential \lambda^{(s)}}\tilde{\lambda}.
\end{equation}
To determine the stability we need to compute the condition for which the fluctuations are suppressed in the dynamics. For this we calculate the eigenvalues of the Jacobian for the system of equations
\begin{align}
    \label{eq:stability_jacobian_eigenvalues}
    k_\pm=-\frac{\varGamma}{2}\pm\sqrt{-\delta^2+\frac{2g\delta}{L^d}\frac{\differential \langle\hat{\Delta}\rangle^{(s)}}{\differential \lambda^{(s)}}}.
\end{align}
Stable solutions must satisfy $\Re{k_\pm}\!<\!0$, the perturbations need to decay back to the stationary state. For $\delta\!\geq\!0$, this gives the condition
\begin{align}
    \label{eq:condition}
    \frac{\delta^2+(\varGamma/2)^2}{2\delta g}>&\frac{1}{L^dZ}\sum_n \text{e}^{-\beta E_n}\left[\frac{\partial \Delta_{nn}}{\partial \lambda}-\Delta_{nn}\left(\frac{\partial\beta}{\partial \lambda}E_n+\beta\frac{\partial E_{n}}{\partial \lambda}\right)\right]\\
    &+\frac{1}{L^{2d}Z^2}\sum_{n,m}\text{e}^{-\beta(E_n+E_m)}\Delta_{nn}\left(\frac{\partial\beta}{\partial \lambda}E_m+\beta\frac{\partial E_{m}}{\partial \lambda}\right)\nonumber
\end{align}
This condition determined the stable character of the thermal steady states under fluctuations of the cavity field.
In practice, due to their complexity, we numerically evaluate the condition in Eq.~(\ref{eq:condition}) for the solutions of Eqs.~(\ref{eq:system_of_equations_1D}) for both the small system exact-diagonalization and in perturbative approach in the thermodynamic limit. 

In order to obtain robust numerical results we evaluate the derivatives using a five-point finite differences scheme with a small step size $\epsilon$
\begin{equation}
    \label{eq:five-point-derivative}
    f'(\lambda)=\frac{-f(\lambda+2\epsilon)+8f(\lambda+\epsilon)-8f(\lambda-\epsilon)+f(\lambda-2\epsilon)}{12\epsilon}.
\end{equation}

\setcounter{equation}{0}
\renewcommand{\theequation}{F.\arabic{equation}}

\subsection{Appendix F: Emergence of bistable solutions}
\label{sec:emergence_of_bistable_solutions}

In this section, we aim to find for some parameter regimes an approximate analytical solution for $(\mu,\beta,\lambda)$ around the onset of the bistability $g_{\text{bi},1}$ in the strong interaction regime satisfying Eqs.~(\ref{eq:system_of_equations_1D}).
We aim to find an expression for the cavity field $\lambda(U,g,\delta,\varGamma)$ and show which terms in the equations are crucial for the emergence of the bistable solutions.  
As discussed in the main text, for Eqs.~(\ref{eq:system_of_equations_1D}) there exists a regime with two stable solutions, $\lambda_{1(2)}$ with lower (higher) cavity field, respectively.
In the following, we derive an approximate condition for $g_{bi,1}$ in terms of the parameters of the model, which needs to be satisfied for the second solution $\lambda_2(U,g,\delta,\varGamma)$ to exist.

\begin{figure}
    \centering
    \includegraphics[width=0.66\textwidth]{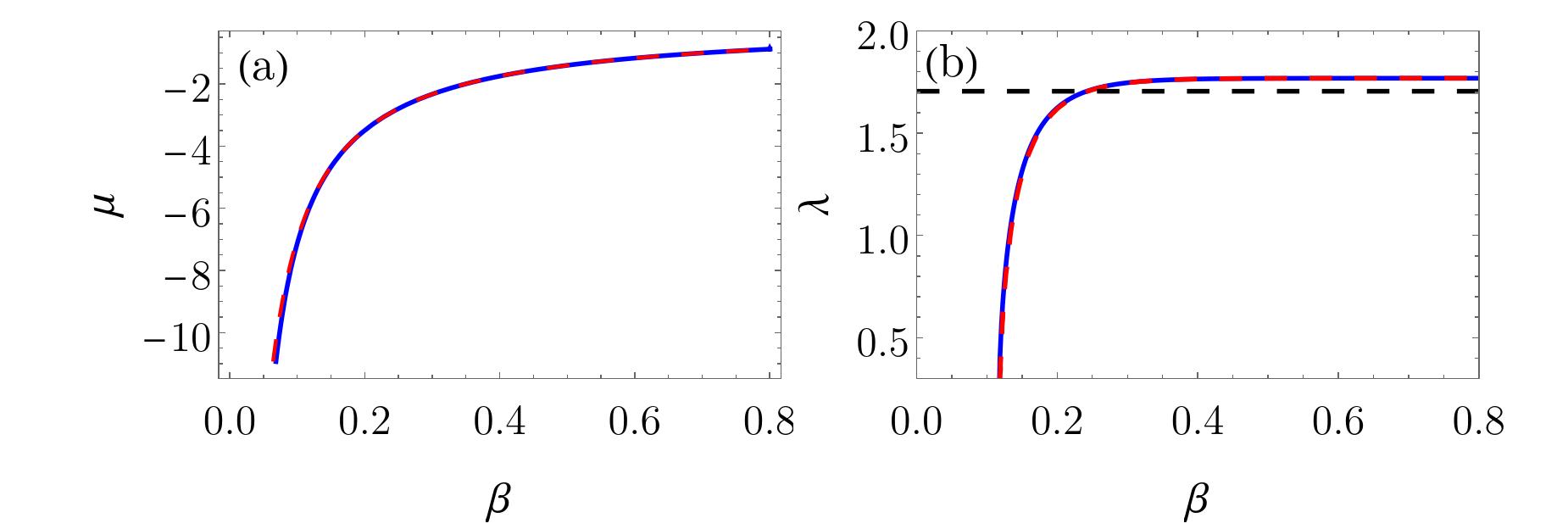}
    \caption{(a) $\mu(\beta)$ approximations, Eq.~(\ref{eq:mu(beta)}) (blue) with Eq.~(\ref{eq:musimple(beta)}) (red dashed) on top 
    (b) $\beta(\lambda)$, full numerical result from particle-conservation and self-consistency Eqs.~(\ref{eq:pc_1D},\ref{eq:sc_1D}) (blue) with approximation Eq.~(\ref{eq:beta(lambda)}) (red dashed) on top, Eq.~(\ref{eq:lambda_0_onset}) (black dashed) at $U/J\!=\!40$, $\hbar\delta/J\!=\!5$, $\hbar\varGamma/J\!=\!3$, $\hbar g/J\!=\!9.65$}
    \label{fig:onset_mu_beta_EOM_approx}
\end{figure}

In the self-organized regime for values of the atom-cavity coupling around the onset of the $\lambda_2$ solution, $g\!\sim\!g_{\text{bi},1}$, we observe that the chemical potential, which determines the average local atomic density $\langle\hat{n}\rangle$ 
[Eq.~(\ref{eq:pc_1D})], only very weakly depends on the value of the cavity field $\lambda$. Thus, we approximate the chemical potential by its value at $\lambda\!=\!0$ in Eq.~(\ref{eq:pc_1D}). 
Solving for $\mu$ this yields the analytical expression
\begin{equation}
    \label{eq:mu(beta)}
    \mu(\beta)=\frac{1}{\beta}\log\bigg[\frac{\!-(\langle\hat{n}\rangle\!-\!1)e^{\beta U}\!+\!\big((\langle\hat{n}\rangle\!-\!1)^2e^{2\beta U}\!-\!(\langle\hat{n}\rangle-2)\langle\hat{n}\rangle e^{\beta U}\big)^{1/2}}{\langle\hat{n}\rangle\!-\!2}\bigg].
\end{equation}
This expression can be simplified considerably for the parameter regime considered in the main text, the value of the filling $\langle\hat{n}\rangle\!=\!1/2$ and assuming the strongly interacting regime $\exp(\beta U)\!\gg\!1$, giving
\begin{equation}
    \label{eq:musimple(beta)}
    \mu_\text{simple}(\beta)=-\log(2)/\beta.
\end{equation}
In Fig.~\ref{fig:onset_mu_beta_EOM_approx}(a) we show the very good agreement between the two expressions for $U/J=40$. 
In this approximation, we obtain that the chemical potential is proportionally increasing with the temperature. 

In the full solutions of the equations of motion we observed that around $g_{\text{bi},1}$ for $\lambda_2$ we have a rather low corresponding temperature [see main text Fig.~2, reasonable agreement with zero-temperature mean-field method]. Thus, as we are interested only in the terms appearing in Eqs.~(\ref{eq:system_of_equations_1D}) with the largest contribution, we neglect in the following the states characterized by the highest energies.
For the regime we are considering these are states with high unit-cell occupation, with either $\geq\!3$ particles per unit cell, or $2$ particles on the high potential sublattice. We refer to this in the following as the \emph{low-energy approximation}, denoted by the index "le".

Within the low energy approximation and using the simplified expression Eq.~(\ref{eq:musimple(beta)}) for the chemical potential $\mu$ the self-consistency condition simplifies considerably to the relation
\begin{equation}
    \label{eq:selfconsistency_approx}
    \lambda=\frac{\delta g}{(\varGamma/2)^2+\delta^2}\tanh(\beta g\lambda/2).
\end{equation}

Even though this relation is not analytically solvable for $\lambda$, one can obtain an expression for the temperature
\begin{align}
    \label{eq:beta(lambda)}
    \beta(\lambda)&=\frac{1}{ g\lambda}\log\left[\frac{\left((\varGamma/2)^2+\delta^2\right)\lambda+\delta g}{-\left((\varGamma/2)^2+\delta^2\right)\lambda+\delta g}\right]
\end{align}
within this approximation. This relation is shown in Fig.~\ref{fig:onset_mu_beta_EOM_approx}(b) and compared to the numerical solution of the full self-consistency equation beyond the low energy approximation.
A good agreement is found. In particular, the value of $\lambda$ rises extremely rapidly a low inverse temperatures and then bends to saturate at larger values of the inverse temperatures corresponding to the low temperature limit to the value of
$\lambda_{\textrm max}\!=\!\delta g/((\varGamma/2)^2+\delta^2)$. We find that the second solution which gives the onset of the bistability and thus, also satisfies the equation of motions lies close to the maximal bending of the curve which lies close to the where $\beta g\lambda\!=\!4$. We will use this point later as an expansion point. 

Using Eqs.~(\ref{eq:musimple(beta)})-(\ref{eq:beta(lambda)}), we rewrite the energy transfer Eq.~(\ref{eq:EOM_1D_terms}) as a function of $\lambda$. 

\begin{figure}
    \centering
    \includegraphics[width=\textwidth]{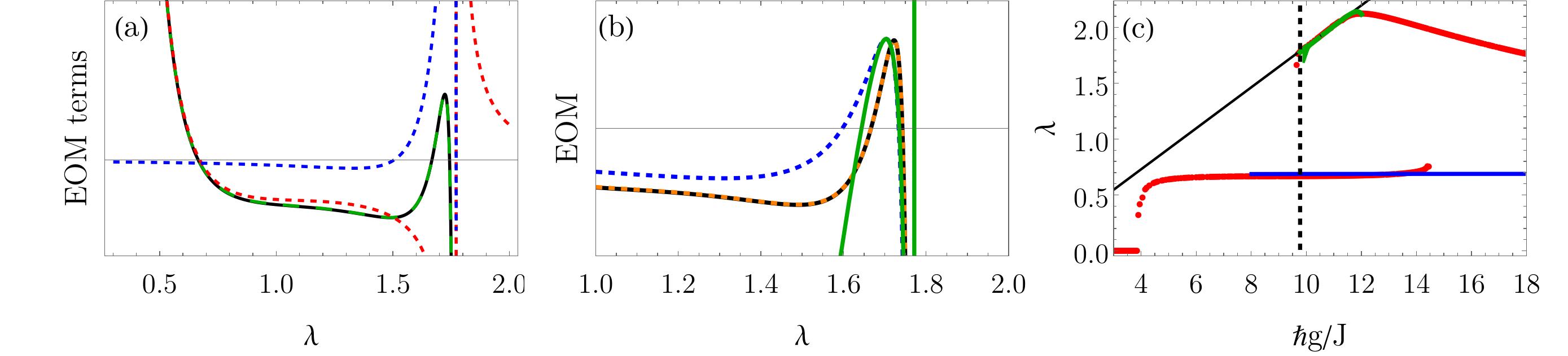}
    \caption{(a) terms in EOM Eq.~(\ref{eq:EOM_1D_terms}) using Eqs.~(\ref{eq:musimple(beta)}),(\ref{eq:beta(lambda)}) all terms (black, dashed), lines 1+2 (green), line 1 (red), line 2 (blue) 
    (b) $\text{EOM}_\text{le}[\mu_\text{simple}(\beta(\lambda)),\beta[\lambda],\lambda]$ [Eq.~(\ref{eq:EOM_1D_terms})] (black), neglect terms $\propto\!1/(2 g\lambda\!+\!U/\hbar)$ (orange), $\text{EOM}_\text{le, approx}$ simplified to Eq.~(\ref{eq:EOM_approx}) (blue), Padé-approximation of Eq.~(\ref{eq:EOM_approx}) (green) 
    (c) full numerical solution for $\lambda$ (red), $\lambda_\text{MF}\!=\!\lambda(\Delta\!=\!N/L^d)$ [Eq.~(\ref{eq:selfcon})] (black), solution of Padé-approx. of Eq.~(\ref{eq:EOM_approx}) (green), approximate solution $\lambda_{1,\text{approx}}$ [Eq.~(\ref{eq:lambda_0_Uinf})] (blue), vertical dashed line [Eq.~(\ref{eq:condition_gbi1})], 
    at $U/J\!=\!40$, $ \hbar g/J\!=\!9.65$, $\hbar\delta/J\!=\!5$, $\hbar\varGamma/J\!=\!3$}
    \label{fig:EOM_approx}
\end{figure}

Taking a look at the individual terms, we observe that taking only the terms $\propto\!1/(2g\lambda)$ [first line Eq.~(\ref{eq:EOM_1D_terms}), red, dashed line in Fig.~\ref{fig:EOM_approx}(a)] into account, gives very good agreement of zero value of the EOM at low values of $\lambda\!\approx\!0.69$.  
This approximation corresponds to the $U\!\to\!\infty$-limit since we neglect the possibility of double occupied sites and we nicely reproduce the $\lambda_1$ solution [blue line in Fig.~\ref{fig:EOM_approx}(c)], yielding 
\begin{equation}
    \label{eq:lambda_0_Uinf}
    \lambda_ {1 \text{approx}}=\sqrt{\frac{\left((\varGamma/2)^2+\delta^2\right)^2-4\delta^2 g^2}{2\left((\varGamma/2)^2+\delta^2\right)^2-4 g^2\left((\varGamma/2)^2+2\delta^2\right)}}
\end{equation}
However, this does not capture the disappearance of the $\lambda_1$-solution at $g_{\text{bi},2}$ and as it is seen in Fig.~\ref{fig:EOM_approx}(a) the additional two values at which the EOM vanishes around $\lambda\!\approx\!1.6$ which correspond to the unstable solution and the $\lambda_2$-solution cannot be recovered by this simple approximation.  

For the appearance of the $\lambda_2$-solution further contributions $\propto\!1/(2g\lambda\!-\!U/\hbar)$ [second line Eq.~(\ref{eq:EOM_1D_terms}), blue line in Fig.~\ref{fig:EOM_approx}(a)] need to be taken into account. This shows the importance of the interaction in the generation of this bistability. 
Neglecting terms $\propto\!1/(2g\lambda\!+\!U/\hbar)$, Eq.~(\ref{eq:EOM_1D_terms}) simplifies to
\begin{align}
\label{eq:EOM_approx}
    0=&\frac{1}{4 g\lambda}\left[\frac{\big(\!(\varGamma/2)^2\!+\!\delta^2\big)\lambda-\delta g}{\big(\!-\!\big(\!(\varGamma/2)^2\!+\!\delta^2\big)\lambda-\delta g\big)\big(\!(\varGamma/2)^2\!+\!(2g\lambda-\delta)^2\big)}-\frac{-\big(\!(\varGamma/2)^2\!+\!\delta^2\big)\lambda-\delta g}{\big(\!\big(\!(\varGamma/2)^2\!+\!\delta^2\big)\lambda-\delta g\big)\big(\!(\varGamma/2)^2\!+\!\left(2 g\lambda+\delta\right)^2\big)}\right]\\
    &+\frac{1}{4(2 g\lambda\!-\!U/\hbar)}\left[\frac{1}{(\varGamma/2)^2\!+\!\left(2 g\lambda\!-\!U/\hbar \!-\!\delta\right)^2}-\frac{\Big(\frac{-\left(\!(\varGamma/2)^2\!+\!\delta^2\right)\lambda-\delta g}{\left(\!(\varGamma/2)^2\!+\!\delta^2\right)\lambda-\delta g}\Big)^{\frac{2 g\lambda-U/\hbar}{g\lambda}}}{\!(\varGamma/2)^2\!+\!\left(2 g\lambda-U/\hbar+\delta\right)^2}\right]\nonumber
\end{align}
[blue dashed line in Fig.~\ref{fig:EOM_approx}(b)]. 
This equation is still too complicated to be solved for $\lambda_2$ analytically, thus, for the terms that only slowly varying with $\lambda$ we replace $\lambda$ with a constant value empirically determined from Eq.~(\ref{eq:selfconsistency_approx}) [see black dashed line in Fig.~\ref{fig:onset_mu_beta_EOM_approx}(b)].
\begin{equation}
    \label{eq:lambda_0_onset}
    \lambda_0=\frac{\delta g}{(\varGamma/2)^2+\delta^2}\tanh(2)
\end{equation}
Furthermore, we approximate the terms in Eq.~(\ref{eq:EOM_approx}) by rational functions using a Padé-approximation \cite{Baker_Graves-Morris_1996} ${\displaystyle [m/m]_{\text{EOM}}(\lambda)}$ to order $m\!=\!1$ around multiple expansion points $\lambda_0$, depending on the divergence points of the individual terms [green line in Fig.~\ref{fig:EOM_approx}(b)].
For the second term in the first line of Eq.~(\ref{eq:EOM_1D_terms}) we choose the $U\!\to\!\infty$-solution for $\lambda_0$ given in Eq.~(\ref{eq:lambda_0_Uinf}) and for the second line terms the expression for $\lambda_0$ in Eq.~(\ref{eq:lambda_0_onset}).
The region of convergence for the approximation strongly depends on the expansion points, and we found this choice to be working the best. The approximations employed capture well the maximum and following zero-crossing exhibited by the numerically solved full system of equations. [compare black, dashed and green lines in Fig.~\ref{fig:EOM_approx}(b)].

After performing the approximations described above, the resulting equations can be solved analytically. 
We obtain a function for $\lambda_2$ around the onset of the bistability, which nicely reproduces the numerical solution for the full system of equations up to the maximum $\lambda_2$-value [see Fig.~\ref{fig:EOM_approx}(c)].
The approximate analytical solution $\lambda_2(U, g,\delta,\varGamma)$ has a real solution only for 
\begin{equation}
    \label{eq:condition_gbi1}
    g\ge\sqrt{\frac{(U/\hbar-\delta)}{2\delta}\left((\varGamma/2)^2+\delta^2\right)}
\end{equation}
This gives an estimate for $g_{\text{bi},1}$ [main text Eq.~(5)]. Comparing to the numerically determined onset of the bistability region we see a nice agreement in the parameter regime of large interaction strengths $U/J$, low values for the detuning $\hbar\delta/J$ and dissipation strengths $\hbar\varGamma/J$ [see main text Fig.~3].

\end{document}